\documentclass[twocolumn]{aastex631}

\usepackage{gensymb}

\received{March 29, 2023}
\accepted{April 26, 2023}

\publishjournal{PASP}

\begin{document}

\title{Scientific CMOS sensors in Astronomy: IMX455 and IMX411}

\correspondingauthor{Miguel R. Alarcon}
\email{mra@iac.es}

\author[0000-0002-8134-2592]{Miguel R. Alarcon}
\affiliation{Instituto de Astrofísica de Canarias (IAC), C/ Vía Láctea, s/n, E-38205, La Laguna, Spain}
\affiliation{Departamento de Astrofísica, Universidad de La Laguna (ULL), E-38206 La Laguna, Canarias, Spain}

\author[0000-0002-9214-337X]{Javier Licandro}
\affiliation{Instituto de Astrofísica de Canarias (IAC), C/ Vía Láctea, s/n, E-38205, La Laguna, Spain}
\affiliation{Departamento de Astrofísica, Universidad de La Laguna (ULL), E-38206 La Laguna, Canarias, Spain}

\author[0000-0002-2394-0711]{Miquel Serra-Ricart}
\affiliation{Instituto de Astrofísica de Canarias (IAC), C/ Vía Láctea, s/n, E-38205, La Laguna, Spain}
\affiliation{Departamento de Astrofísica, Universidad de La Laguna (ULL), E-38206 La Laguna, Canarias, Spain}

\author[0000-0001-9502-781X]{Enrique Joven}
\affiliation{Instituto de Astrofísica de Canarias (IAC), C/ Vía Láctea, s/n, E-38205, La Laguna, Spain}

\author[0000-0003-4009-2061]{Vicens Gaitan}
\affiliation{Aplicaciones en Informática Avanzada (AIA), E-08172 Sant Cugat del Vallès, Catalonia, Spain}

\author{Rebeca de Sousa}
\affiliation{Departamento de Historia y Filosofía de la Ciencia, la Educación y el Lenguaje, Universidad de La Laguna (ULL), E-38206 La Laguna, Canarias, Spain}




\begin{abstract}

Scientific complementary metal-oxide-semiconductor (CMOS) detectors have developed quickly in recent years thanks to their low cost and high availability. They also have some advantages over charge-coupled devices (CCDs), such as high frame rate or typically lower readout noise. These sensors started to be used in astronomy following the development of the first back-illuminated models. Therefore, it is worth studying their characteristics, advantages, and weaknesses. One of the most widespread CMOS sensors are those from the Sony IMX series, which are included in large astronomical survey projects based on small and fast telescopes because of their low cost, and capability for wide-field and high-cadence surveys. In this paper, we aim to characterize the IMX455M and IMX411M sensors, which are integrated into the QHY600 and QHY411 cameras, respectively, for use in astronomical observations. These are large (36 $\times$ 24 and 54 $\times$ 40 mm) native 16 bit sensors with 3.76 $\mu$m pixels and are sensitive in the optical range. We present the results of the laboratory characterization of both cameras. They showed a very low dark current of 0.011 and 0.007 e$^{-}$ px$^{-1}$ s$^{-1}$ @--10\degree C for the QHY600 and QHY411 cameras, respectively. They also show the presence of warm pixels, $\sim$0.024\% in the QHY600 and 0.005\% in the QHY411. Warm pixels proved to be stable and linear with exposure time, and are therefore easily corrected using dark frames. Pixels affected by the Salt \& Pepper noise are $\sim$2\% of the total and a method to correct for this effect is presented. Both cameras were attached to night telescopes and several on-sky tests were performed to prove their capabilities. On-sky tests demonstrate that these CMOS behave as well as CCDs of similar characteristics and (for example) they can attain photometric accuracies of a few milli-magnitudes.

\end{abstract}



\section{Introduction}
Historically, observational astronomy has been determined by the development of imaging technology. In the early twentieth century, the introduction of the photographic plate marked an unprecedented revolution, but their very limited quantum efficiency (QE) of $<$0.1\% made it imperative to develop more effective alternatives. The turning point came in 1969 with the appearance of charge-coupled devices (CCDs, \cite{boyle1970charge}), which are silicon-based sensors with
two-dimensional arrays of photosensitive units (pixels). These
devices produced improved linearity and quantum efficiency
over photographic plates and simplified data analysis thanks to
their analog-to-digital conversion (ADC). The first astronomical
image that was taken with a CCD appeared in 1975 when
scientists from NASA’s Jet Propulsion Laboratory obtained an
image of the planet Uranus at a wavelength of 8900 \r{A} \citep{janesick1987sky}. The development and refinement of CCDs have
increasingly made these sensors the most widely used option
for the manufacture of astronomical instruments.

Over the last two decades, the rise of alternative technologies
has undermined the prevalence of CCD sensors.
Complementary metal-oxide-semiconductor (CMOS) image
sensors started to be developed in the 1990s \citep{Fossum1997CMOSAP}. However, they still had a number of disadvantages when
compared with CCDs, such as lower dynamic range (DR), and
poorer linearity and sensitivity \citep{bigas2006review}. Although
CMOS sensors soon established themselves in the consumer
market, their inherent constraints restricted their application in
certain fields, especially those related to science. To overcome
the typical limitations of CMOS, the so-called scientific CMOS
(sCMOS) were introduced in 2009 as a result of a collaboration
between Andor Technology, Fairchild Imaging (BAE Systems)
and PCO Imaging \citep{breakthrough2009scmos}. This new generation
of sensors combined high frame rates, reasonable pixel and
sensor sizes, and quantum efficiencies comparable to
CCDs---especially back-illuminated (BI) sCMOS \citep{bi_scmos}---and a considerable reduction in the noise
levels that are traditionally associated with CMOS.

\begin{figure*}
    \includegraphics[width=\textwidth]{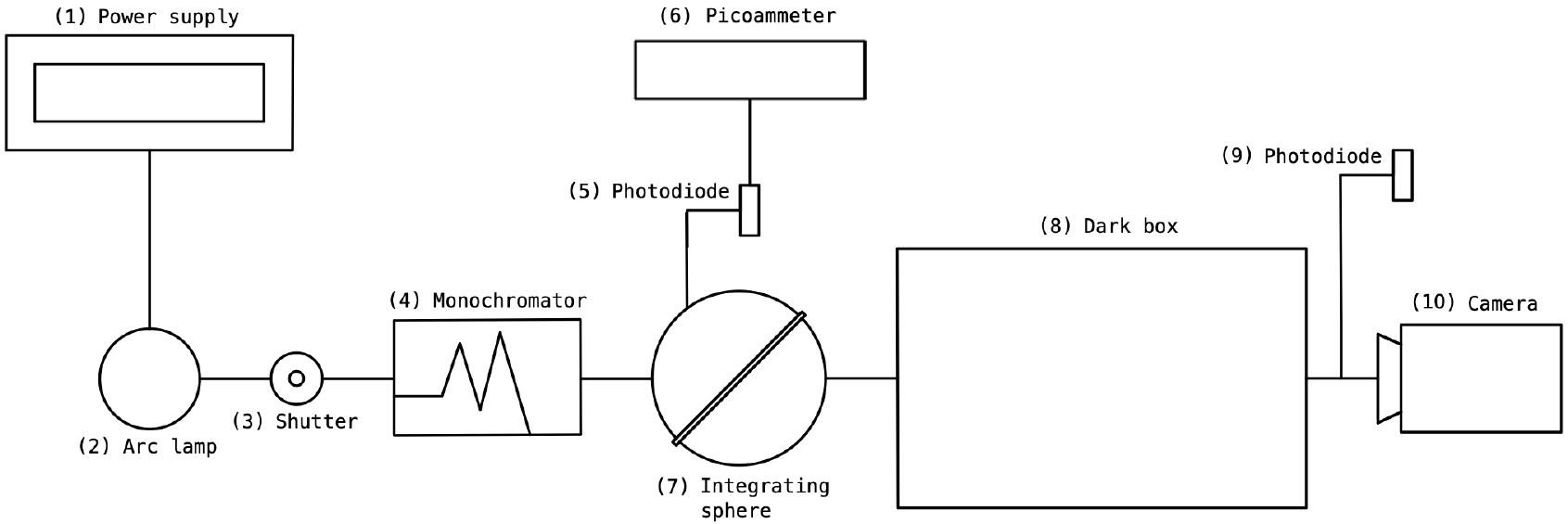}
    \caption{Diagram of the optical test bench set-up, with the (1) Newport 68945 digital power supply, (2) Newport M-66881 QTH lamp, (3) Newport 76994 shutter, (4)
Newport Oriel Cornerstone monochromator, (5) Hamamatsu S1336-5B1 photodiode, (6) Labsphere SC6000, (7) Labsphere US-080-SF/SL integrating sphere, (9)
Hamamatsu S2281 photodiode, and (10) QHY600M Pro / QHY411M}
    \label{fig:lab}
\end{figure*}

Scientific CMOS are beginning to be accepted in astronomy. The first analyses in this respect mainly involve two sensors: CIS2051 (later rebranded as CIS2521), which was developed by Andor Technology and integrated into the Neo series detectors; and GSENSE, which was developed by Gpixel and integrated by several companies, e.g., Finger Lakes Instrumentation, QHYCCD, and Andor Technology. Experiments performed with Andor Neo showed low readout noise ($\sim$1 e$^{-}$) and linearity deviations up to the saturation point within the expected range of $\pm$1\%. However, two features were noticed: the transfer curve was observed to bend, even at low signal levels, which was possibly caused by the non-linearity of an amplifier stage; and the signal shows certain irregularities with high variance in the transition region between high and low gain modes \citep{schildknecht2013improved}. This, along with a limited QE due to being front-illuminated and a lower fill factor \citep{qiu2013evaluation}, means that this sensor is not the most suitable for regular observations, and its usefulness is mostly constrained to bright objects and high frame rates. The back-illuminated GSENSE2020 sensor, integrated in the Andor Marana, was recently tested for its application in astronomy by \cite{qiu2021research} and \cite{Karpov2021}. This sensor showed a low readout noise (1.6 e$^{-}$), good linearity ($99.7\%$), and a stable bias. This sensor family has a dual-amplifier structure in which two 12 bit images are taken simultaneously and are then merged into a single 16 bit high dynamic range image. However, this mechanism is susceptible to jumps and instabilities around the transition region between high and low gain. In addition, edge glowing and charge persistence over long periods of time is also present. The capabilities of these sensors decrease considerably with exposure times longer than several seconds, which is a consequence of these effects and the increase of dark current; however, they are still suitable instruments for high frame rate observations.

Many leading manufacturers are now developing instrumentation based on next-generation sCMOS sensors, but their suitability for general use in astronomy is still largely unexplored. One of the most widespread sensors are those from the Sony IMX series, which are being included in large projects such as the Argus Optical Array \citep{argus2022}, Large Array Survey Telescope \citep{Ofek2023} or the next generation of telescopes for the ATLAS project \citep{Tonry2018} , such as the one that will be installed at Teide Observatory, ATLAS-Teide \citep{Licandro2023}, because of their low cost and capability for wide-field and high-cadence surveys.

In this paper, we present the results of the laboratory characterization of the sCMOS BI Sony IMX455M and IMX411M sensors when integrated into the QHY600M and QHY411M cameras. In Section \ref{sec:methods}, the two devices are tested, and the laboratory set-up and the telescopes that we used are presented. Sections \ref{sec:bias} and \ref{sec:dark} describe the tests performed under dark conditions: spatio-temporal variation of bias and dark and contaminating effects, such as random telegraph noise and the presence of warm pixels. In Section \ref{sec:PTC}, the main operating features such as the gain fix pattern noise and linearity are verified using the photon transfer curve. The quantum efficiency measures are shown in Section \ref{sec:QE} and the charge persistence effect reported in other sCMOS sensors is reviewed in Section \ref{sec:persistence}. Finally, several approaches to processing telescope data and on-sky results based on this analysis are discussed in Section \ref{sec:discussion}.

\begin{deluxetable*}{lcc}
\tablecaption{Technical data of QHY600M Pro and QHY411M sCMOS cameras provided by the manufacturer.}\label{tab: QHY600/QHY411} 
\tablehead{
\colhead{Feature} & \colhead{QHY600M Pro} & \colhead{QHY411M}}
\startdata 
Sensor & Sony BI IMX455M & Sony BI IMX411M\\\\
Sensor size \scriptsize{(diagonal)} & 43.3 mm & 66.7 mm\\\\
Pixel size & \multicolumn{2}{c}{3.76 $\times$ 3.76 $\mu$m}\\\\
Pixel area & 9600 $\times$ 6422 & 14304 $\times$ 10748\\\\
Effective pixels & 61.17 Mpx & 151 Mpx\\\\
Max full frame rate \\ \scriptsize{(USB 3.0 port, full frame} \\ \scriptsize{and 16-bit output)} & 2.5 fps & 1 fps\\\\
A/D sample depth \\ \scriptsize{(1$\times$1 binning)} & \multicolumn{2}{c}{16--bit}\\
Shutter type & \multicolumn{2}{c}{Rolling shutter}\\\\
Cooling system \\ \scriptsize{(temperatures below ambient)} & Air cooling \scriptsize{(-30C)} & 
Air cooling \scriptsize{(-35C)} \\ & & 
Water cooling \scriptsize{(-45C)}
\enddata
\end{deluxetable*}

\section{Methods}\label{sec:methods}
\subsection{Instruments}
The QHY600M\footnote{https://www.qhyccd.com/scientific-camera-qhy600pro-imx455/} camera is based on the back-illuminated IMX455 monochrome sensor that is manufactured by SONY,
which is a full-frame (35 mm format) sensor with $9576 \times 6388$, 3.76 $\mu$m square pixels. The QHY411M\footnote{https://www.qhyccd.com/scientific-camera-qhy411-qhy461/} camera is based on
SONY’s IMX411 monochrome sensor, which is also backilluminated
but with a larger sensor size, $14304\times10748$ (equivalent to medium-format cameras, $54 \times 40$ mm). Both of
these sensors include an overscan region of 33 and 91 rows,
respectively, and they are native ADC sampled at 16 bit, which is a
significant change from previous generations of sCMOS sensors
that were based on 12 bit image merging (see \cite{Karpov2021}). The main features of the cameras are listed in Table \ref{tab: QHY600/QHY411}. We used the
QHY600M Pro version, which allows a faster $2 \times 10$ Gbps fiber
connection to a frame grabber with an additional 4 GB of DDR3
memory---although in this paper all tests were done via USB 3.0
connection---and triggering the rolling shutter via an external GPS
with an accuracy of more than a microsecond. It is worth
mentioning that both sensors, especially the IMX455, are used in
cameras from other manufactures, e.g., Atik Apx60, ZWO
ASI6200MM Pro.

These cameras can operate in several modes and gain settings,
which essentially change their gain, readout noise (RON), and
full-well capacity (FWC). In this paper, we have focused on
those that are considered most appropriate for use in astronomy
because they maintain a good balance between these characteristics,
which are Mode\#1 (High Gain Mode) and gain setting 0 (hereafter, \#1@0) on the QHY600M Pro and Mode \#4, and gain setting 0 (hereafter, \#4@0) on the QHY411M, although
some results for other modes are also shown.

\subsection{Optical test bench}
The evaluation of the sensors was performed using existing
experimental equipment, which are available at the Laboratory of
Imaging and Sensors for Astronomy (LISA), at the Instituto de
Astrofísica de Canarias (Tenerife, Spain). The set-up is schematically
shown in Figure \ref{fig:lab}. A Newport M-66881 QTH lamp was
connected to A Newport 68945 digital power supply, with
intensity stabilization. In the Newport Oriel Cornerstone monochromator,
the option to apply an order sorting filter was selected.
This filter blocked higher diffraction orders from interfering with
the selected wavelengths. A Newport 76994 shutter was placed
between the lamp and the monochromator to control the light
beam before taking each dark frame. A Labsphere US-080-SF/SL
integrating sphere, placed next to the monochromator, had incorporated a Hamamatsu S1336-5B1 photodiode, which was in
turn connected to a Keysight B2980A picoammeter. Next, a dark
box was located, followed by the test of the camera. A second
photodiode (Hamamatsu S2281) was connected to the exit port of
the dark box. The conversion factors between the intensity
measured by the picoammeter on the first photodiode and the
radiant power received on the second photodiode at the exit of the
dark box were characterized, in addition to the correction to the
distance between the photodiode and the sensor back focus.
During the tests, the laboratory temperature was about $23\degree$C and the humidity was 40--50\%. Both cameras were air-cooled, with operating temperatures of $-5\degree$C for the QHY600M Pro and $0\degree$C for the QHY411M. 

\subsection{Telescopes}
The sky tests were performed with two of the robotic
telescopes (Telescopios Abiertos Robticos, TAR) of Teide
Observatory (Tenerife, Canary Islands, Spain). The
QHY600M Pro was installed on the prime focus of TAR03,
0.46-m $f$/2.8 C18 reflector telescope on a Planewave L500
altaz mount. An ROI of $4096\times4096$ was used, giving a FOV of $41^{\prime}.1\times41^{\prime}.1$ with $0^{\prime\prime}.6$ px$^{-1}$. The QHY411M was mounted on a Meade LX200-ACF 16$^{\prime\prime}$ 0.406-m $f$/10, on an APM GE-300 Direct Drive equatorial mount, with UV/IR-Cut/L, SDSS g$^{\prime}$, r$^{\prime}$ and i$^{\prime}$ 50 mm filterswhich were manufactured
by Baader Planetarium GmbH. The sensor was trimmed to $40 \times 40$ mm, giving a FOV of $34^{\prime}.2 \times 34^{\prime}.2$ arcmin with 0.40$^{\prime\prime}$ px$^{-1}$. Both cameras were connected via USB 3.0 and operated with
air cooling at $-10\degree$C. After the first tests, the QHY411M was
installed in one of the 80 cm telescopes of the Two-meter
Twin Telescope (TTT) project, which is a 0.80-m $f$/6.85 Ritchey--Chrétien altaz telescope that was manufactured by
ASA Astrosysteme GmbH, with a total FOV of $33^{\prime}.4\times 25^{\prime}.0$  and a plate scale of $0^{\prime\prime}.14$ px$^{-1}$.

\section{Results}
\begin{figure}
  \centering
    \includegraphics[width=\linewidth]{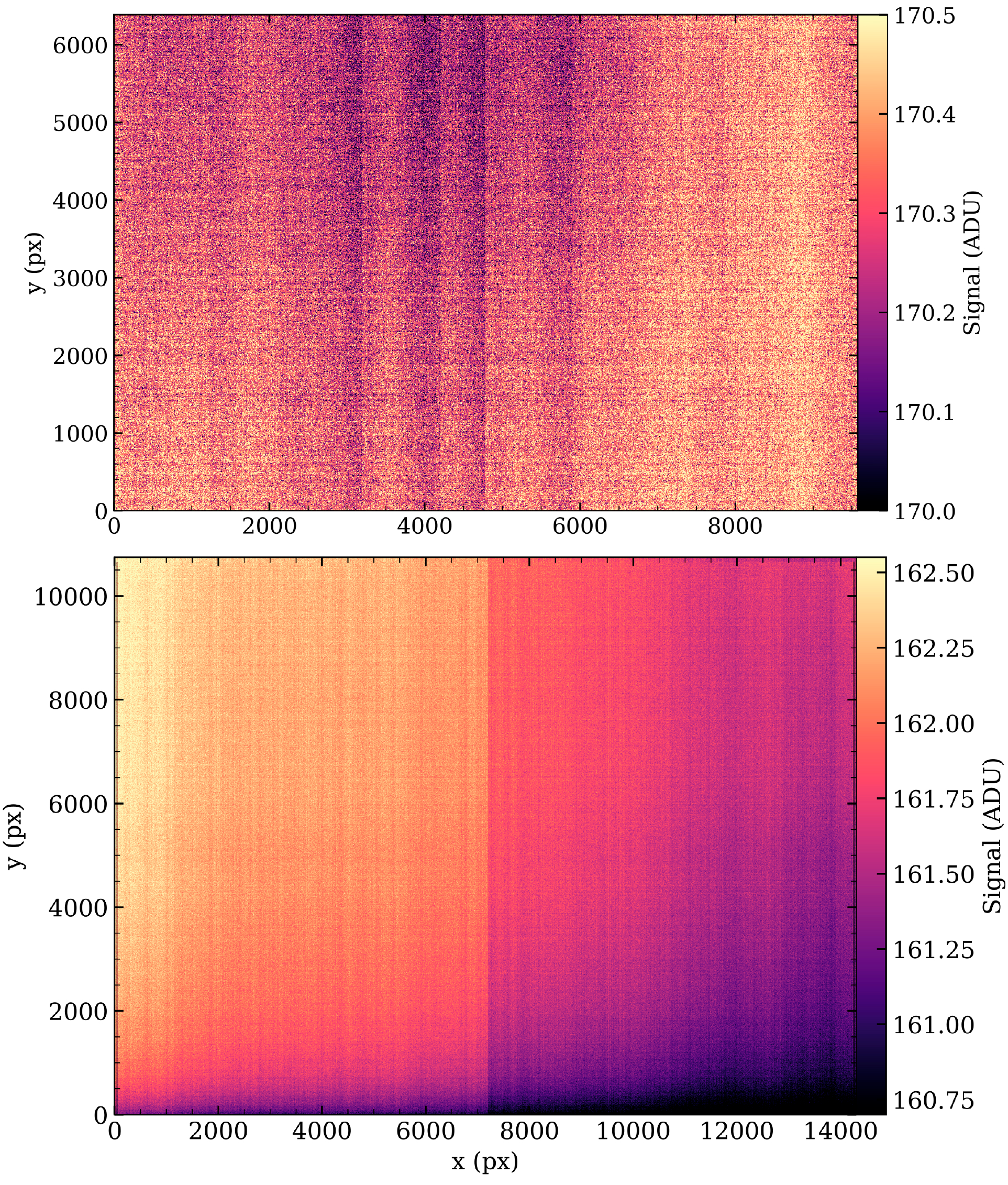}
      \caption{Full-frame master bias of the QHY600M Pro (top) and QHY411M
(bottom) obtained by 3$\sigma$-clipping median stacking of 21 bias frames taken
consecutively.}
        \label{fig:masterbias}
\end{figure}

\begin{figure*}
    \centering
    \includegraphics[width=.49\linewidth]{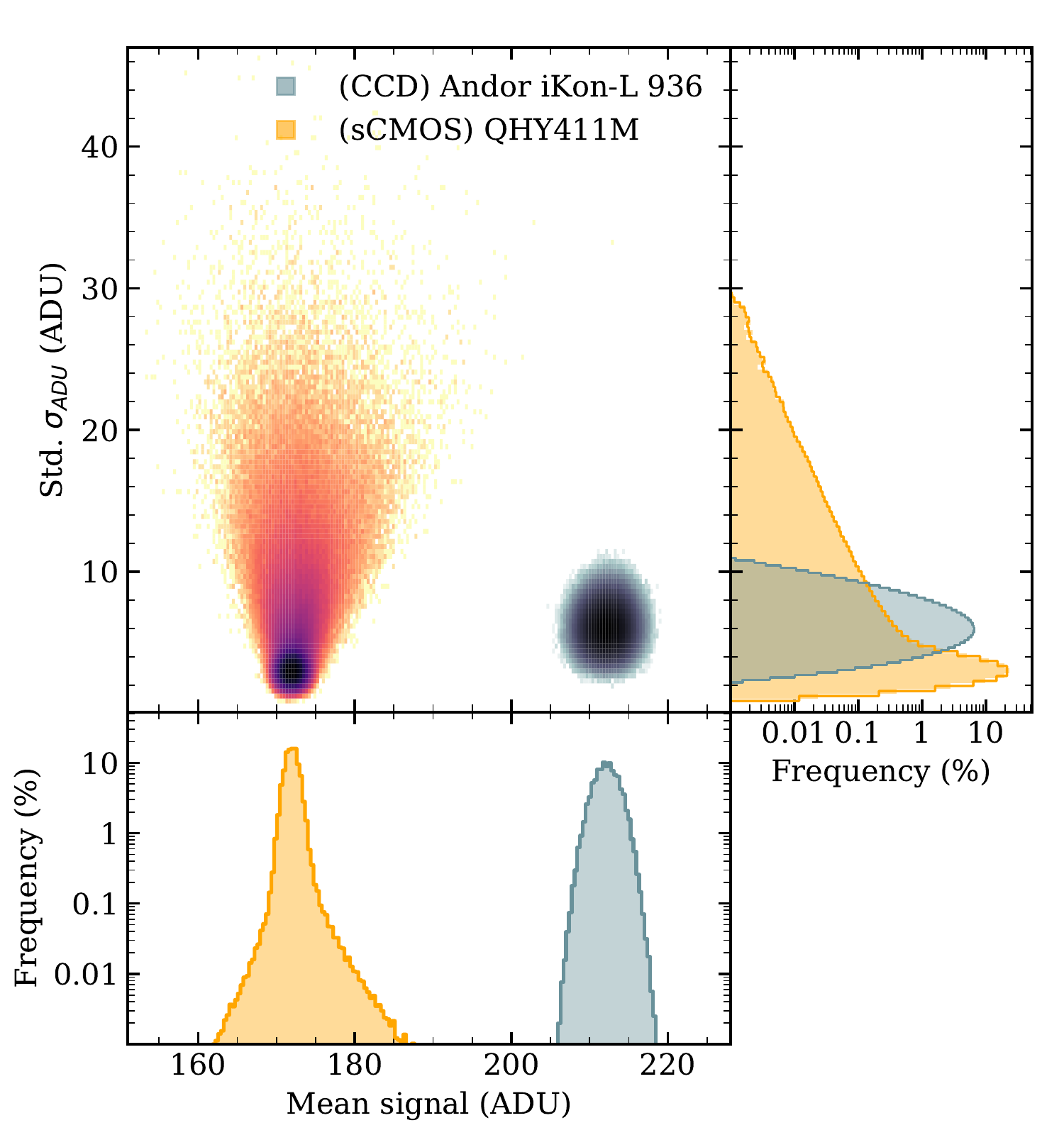}
    \includegraphics[width=.49\linewidth]{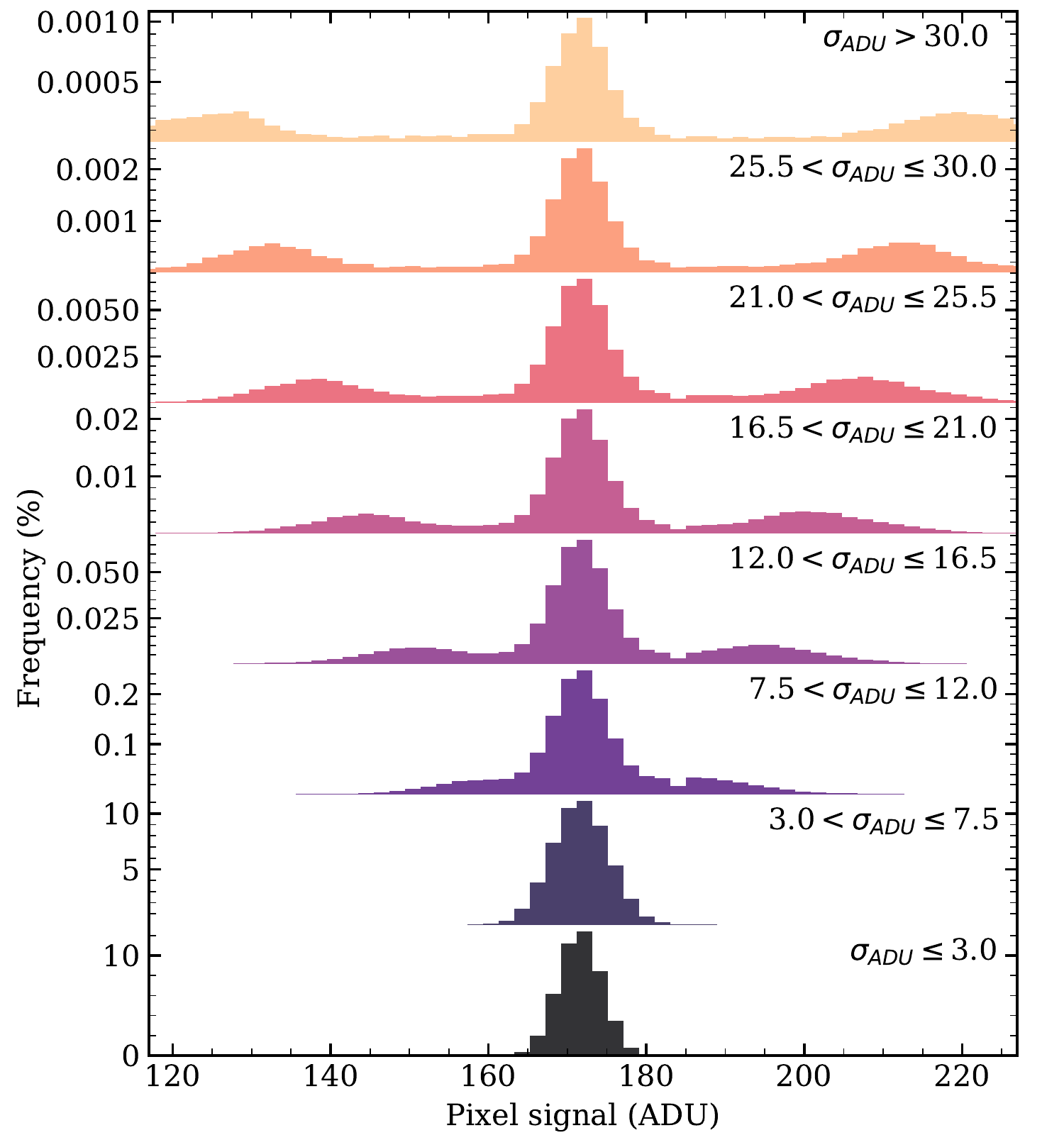}
    \caption{Left-hand panel: Standard deviation vs. mean signal of each QHY411M pixel (warm colors) in 21 consecutive bias frames. Color indicates data density on a
logarithmic scale. The histogram of the distribution of these two variables is included below and to the right-hand side. For comparison, the same information is shown
in blue tones for a CCD, the ANDOR iKon-L 936 BEX2DD. The QHY411M pixels have been grouped according to their standard deviation over the 21 frames and,
for each of them, the signal measured in these frames was taken individually. The distributions of each group are given in the right-hand plot.}\label{fig:RON}
    \end{figure*}

\subsection{Bias stability and "Salt \& Pepper" effect}\label{sec:bias}
The bias frame stability in both cameras was tested in the
laboratory under conditions of complete darkness. In all of the
tests that are presented here, including the telescope runs, the
offset setting was fixed at 10. First, 21 continuous unbinned full
frames bias were obtained and stacked with a 3$\sigma$-clipping
median to obtain the master bias. The result is shown in
Figure \ref{fig:masterbias}. Both cameras show dark signal non-uniformity
(DSNU), with a notable column-to-column pattern and pixel-to-pixel variations. Gradients can be observed in the
QHY600M Pro toward the upper region, with variations of
less than 1 ADU, especially in the QHY411M, with more than
2 ADU of difference between the upper left-hand and the lower right-hand corner.

In the readout process of typical CCDs, the charge is
transferred to one---or several in the case of large sensors---output channels, each with a high-quality amplifier and an
analog-to-digital converter (ADC). In contrast, in CMOS, each
pixel has its own built-in electronics, which contain an
amplifier and ADC. This allows a very high frame rate by
parallelizing the readout process but exposes the sensor to these
kinds of pixel-to-pixel variations, both in darkness (with different readout or thermal noises) and illuminated (with
variations in gain or quantum efficiency). In sCMOS, each
pixel must be understood as an independent sensor, so all
corrections over them, such as subtraction of the bias level,
must be done while bearing in mind that there is no single value
that characterizes all of the pixels simultaneously. For this
reason, although the QHY600M Pro and QHY411M have an
overscan region, it is more accurate to use a full frame master
bias or dark that takes into consideration these non-uniformities
between pixels.

Once the spatial behavior of each pixel in darkness has been
revised, the same 21 bias frames of the QHY411M are used to
see its temporal performance in detail. Now, instead of taking
the three-clipped median, the unclipped standard deviation and
mean signal are plotted against each other, for every single
pixel, in the left-hand plot of Figure \ref{fig:RON}. This shows that most of
the pixels are found in a circular zone with a well-determined
average signal between 170 and 175 ADU, corresponding to
the median bias value, and with a standard deviation below 4
ADU. However, about 11\% of the pixels show a temporal
standard deviation that is higher than this value and also
present deviations in the mean signal, with a kite-shaped
dispersion. If this scatter were caused by pixels exhibiting
higher values of read noise only, then the points would be
distributed vertically, like a gradated column. If, however, the
pixels had a different threshold voltage, i.e., bias level, but
maintained the RON, then the points would be distributed
horizontally. Instead, the kite-shape indicates that either both
occur at the same time for these pixels or that there is an
additional anomalous behavior at play.

For comparison, the same number of frames was taken with
a CCD, the ANDOR iKon-L 936 BEX2-DD\footnote{https://andor.oxinst.com/assets/uploads/
products/andor/documents/andor-ikon-l-936-specifications.pdf}, and the standard
deviation and temporal average of each pixel was also
obtained. The kite-shaped dispersion is not visible in this case.
Furthermore, the histograms along both axes show the expected
Gaussian shape: on the standard axis, a peak centered on the
RON value; and on the mean signal axis, a peak centered on the
average bias level and a width equal to the RON/$\sqrt{21}$. In the
case of the QHY411M---the same behavior is observed in the
QHY600M Pro---the histograms show a narrow peak and tails
toward high standard deviation. This implies that most of the
pixels show a low RON, but some others have a skewed signal
on both sides of the average bias level.

With this distribution, defining a nominal value for RON is
not straightforward. Looking at the standard deviation
histogram, the peak is located at 3.0 ADU (3.1 e$^{-}$), and the
rms at 3.74 ADU (3.83 e$^{-}$), close to the 3.72 e$^{-}$ reported by the manufacturer. From here on, the rms value will
be used as a reference for the RON. For the QHY600M Pro, an
rms value of 3.48 e$^{-}$ was obtained, compared to the 3.67 e$^{-}$ reported by the manufacturer. The values for the other modes
of operation are included in Table \ref{tab:res}.\\

\begin{figure}
  \centering
    \includegraphics[width=\linewidth]{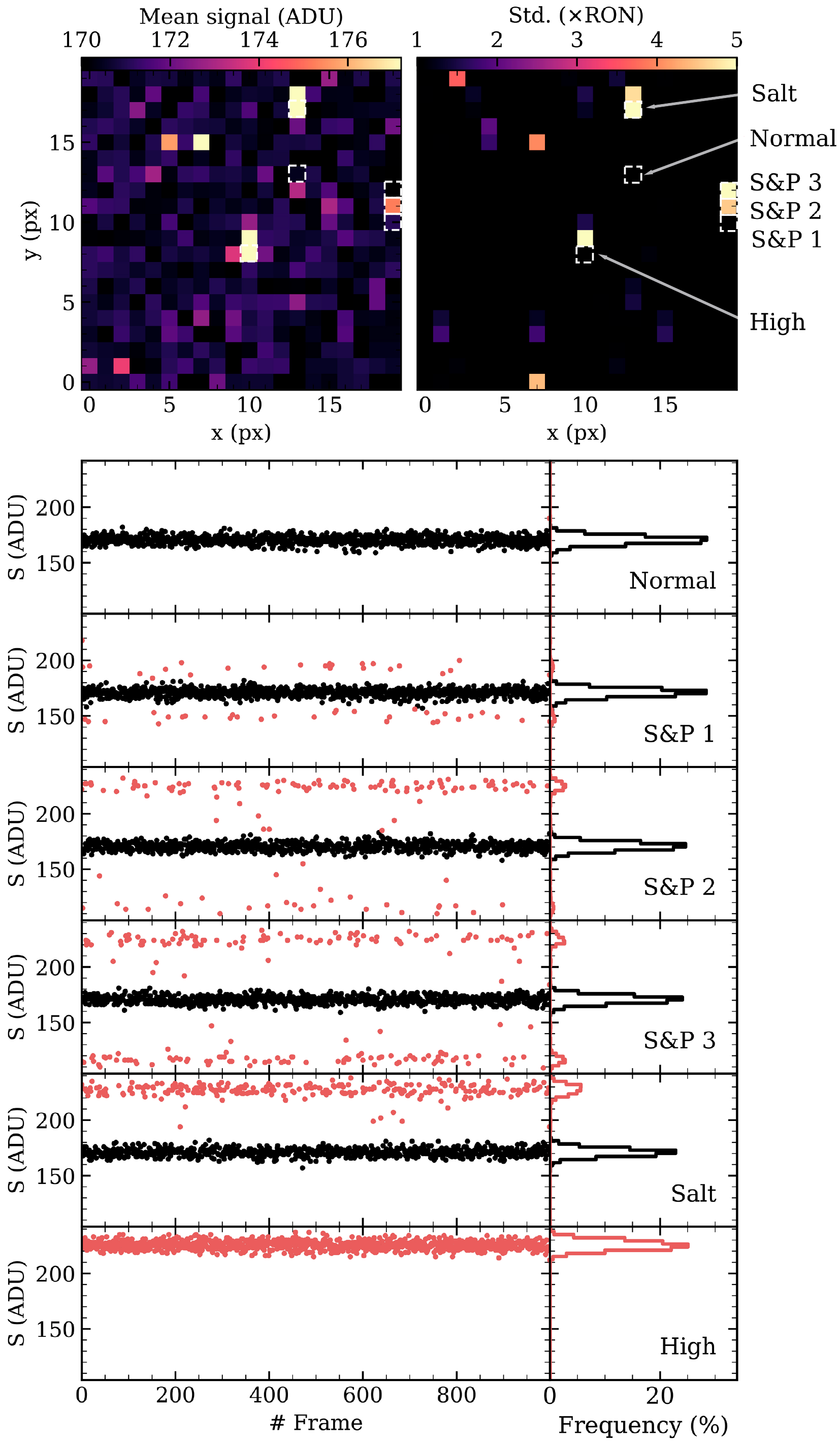}
      \caption{Top: temporal mean signal (left-hand) and standard deviation n terms
of readout noise (right-hand) obtained in a $20 \times 20$ pixel central region of 1000
bias frames taken consecutively with the QHY600M Pro. Bottom: signal vs.
frame number for some relevant pixels identified in the images above. Points
within the average value of the master bias $\pm3\times$ RON are shown in black, with
outliers identified in red. The signal distribution of these pixels is displayed on
the right-hand side.}\label{fig:rtn}
\end{figure}

\begin{figure*}
    \centering
          \includegraphics[width=0.49 \linewidth]{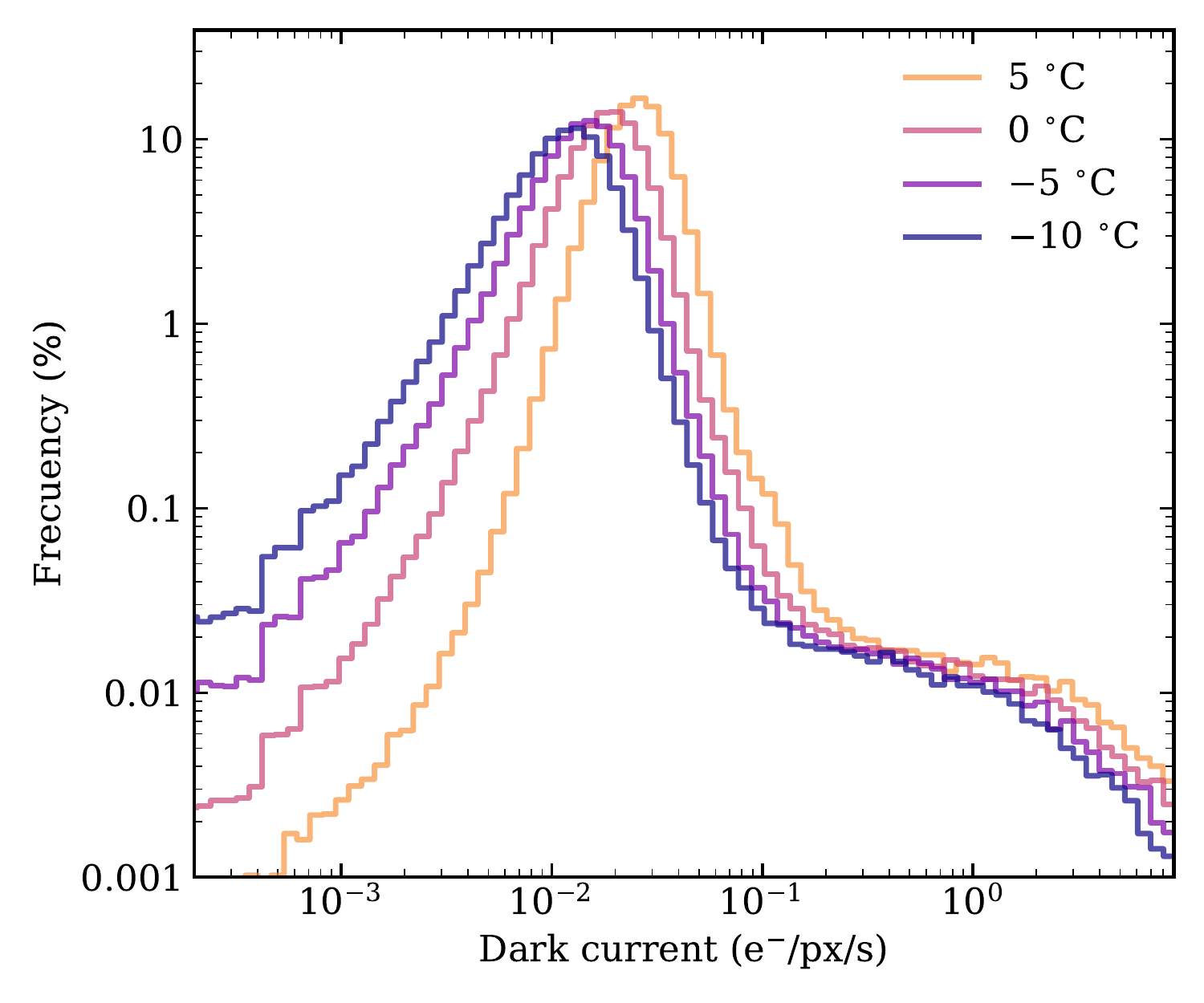}
          \includegraphics[width=0.49 \linewidth]{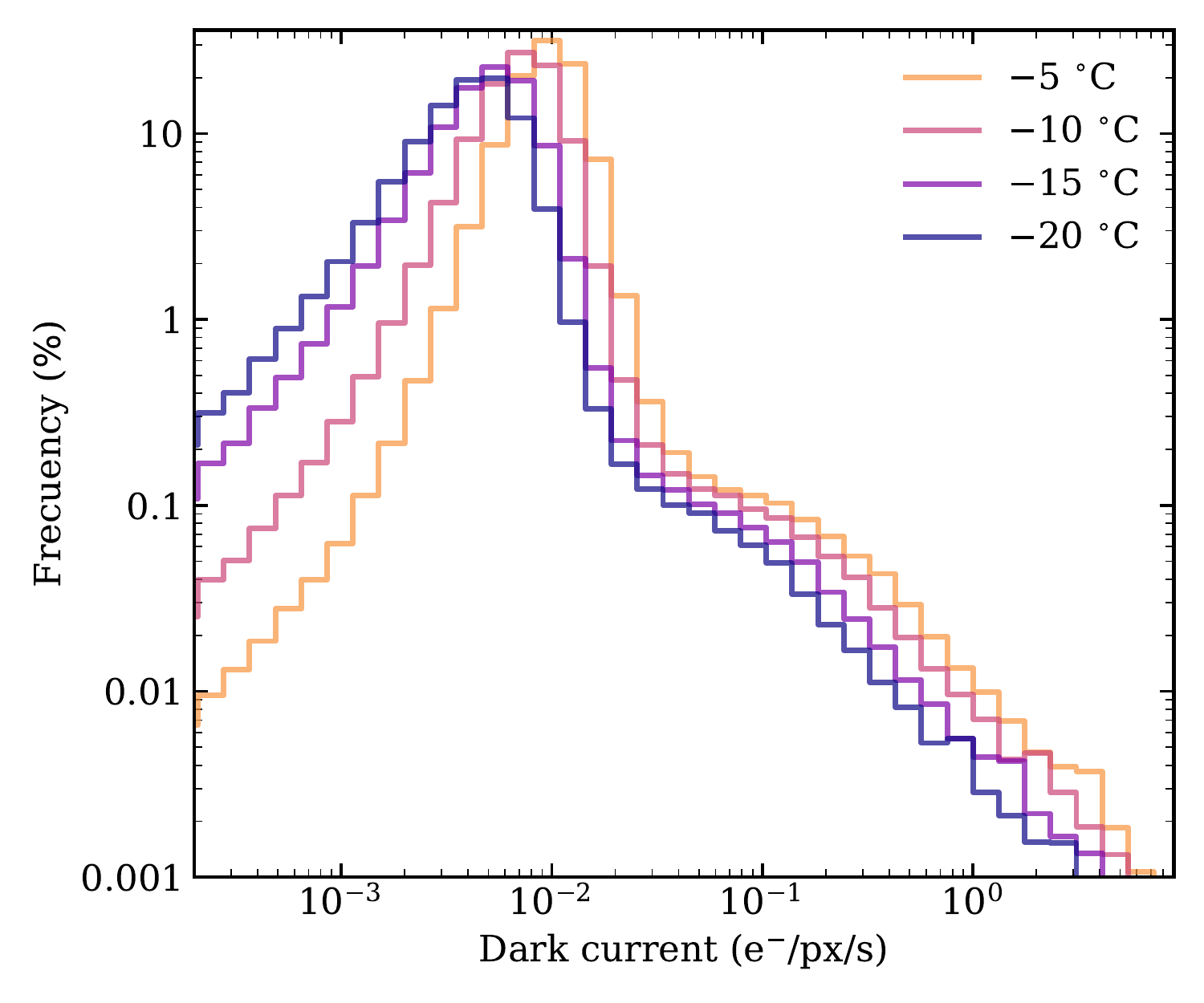} 
    \caption{Dark current distributions for the QHY600M Pro (left-hand panel) and the QHY411M (right-hand panel). They were obtained as the median stacked set of five dark frames with an exposure time of 1000 s.}\label{fig:dark_hist}
\end{figure*}

Returning to the standard versus average plot for the sCMOS
in Figure \ref{fig:dark_hist}, the pixels are now grouped in ranges of std. $\sigma_{\text{ADU}}$ and the signal of each obtained in all of the 21 frames is
retrieved. The distribution of each of these groups is shown in
the right-hand plot of Figure \ref{fig:dark_hist}. For pixels showing a temporal
deviation of the order of the RON ($\sigma_{\text{ADU}}\leq$7.5), the
distribution of the measured signal is Gaussian, similar to that
of a conventional CCD. The large majority of the pixels are in
these first two ranges (45\% of the pixels with $\sigma_{\text{ADU}}\leq$3 and 98\% with $\sigma_{\text{ADU}}\leq$7.5). For the other ranges (with larger
deviations), the pixels begin to show a triple Gaussian
distribution, with two smaller peaks that are centered on both
sides of the central peak and with similar widths. For larger
values of $\sigma_{\text{ADU}}$ the separation of the peaks is greater. In these
anomalous pixels, the pixel distribution is indicative of the
following behavior: while most of the time these pixels return
values that are around the mean value of the bias level with a
normal noise equivalent to readout, their returned values
occasionally jump toward larger or smaller values with defined
separations and different probabilities for each pixel.

The symmetry in the three Gaussian distributions does not
necessarily mean that all of the anomalous pixels jump between
higher and lower levels with the same probability. In that case,
the standard versus average plot would not show a kite-shaped
scatter but a column, given that the average of the signal would
always lie in the central range. To show a kite-shaped
dispersion, there must be some pixels that tend to exhibit more
deviations toward one of the Gaussians, either the upper or the
lower Gaussian, thus skewing the average to either side and
resulting in the widening of the tails of the mean signal
distribution, which is shown in yellow in the lower left-hand
plot of Figure \ref{fig:RON}.\\

For a more detailed insight into the behavior of these pixels
with high standard deviation, 1000 consecutive bias frames
were obtained with the QHY600M Pro. The temporal average
and standard deviation of all the frames has been taken in a
small central region of $20 \times 20$, which is shown in the
upper part of Figure \ref{fig:rtn}. Most of the pixels have an average
signal around 171 ADU and a dispersion below the RON, as
expected. However, some of them show anomalous patterns in
the average value, the standard deviation, or both at the same
time. Several pixels have been selected as samples, showing the
time evolution of the signal in the lower plots. As a reference, a
pixel with normal values of average signal and deviation has
been taken, following a normal distribution over the 1000
frames, with mean 170.4 ADU and standard deviation 3.69 ADU. The pixel labeled S\&P 1 also has an average signal
similar to the others, with a slightly higher standard deviation
of 6.42 ADU but still close to the RON obtained in the previous
section. However, the temporal distribution reveals how some
points, 5.5\% of the total, appear both above and below the
average signal, with a gap of about 20 ADU, which is more
than three times the RON (see the red dots in the plots below).
Other pixels show the same effect more often, e.g., S\&P 3, where 19.3\% of the 1000 frames show an anomalous value of $\pm60$ ADU. This is revealed by a high standard deviation of
23.6 ADU. Although this can be understood as a high RON in
its electronics, it should be noted that its distribution is not a
wide Gaussian but a set of three normal distributions---the main
distribution is centered on 171 ADU, and two smaller ones
corresponding to these random leaps to higher and lower values
at 110 and 230 ADU. The S\&P 2 pixel shows jumps at the
same levels but with a higher proportion of values in the upper
level (10.6\% of the total) than in the lower level, 3\%. Consequently, the signal that was obtained in the average
frame, 174.5 ADU, deviates from the other pixels, as shown in
the image.\\

These jumps between above- and below-average signal
values are observed when blinking between images, regardless
of their exposure time and temperature. A pattern of bright and
dark pixels that appear and disappear from one frame to the
next is observed. This effect, which is sometimes referred as
Salt \& Pepper, is random telegraph noise (RTN). RTN is the
fluctuation of the signal between discrete levels as a
consequence of the capture and emission of charges by defects
or traps located very close to the Si--SiO$_2$ interface \citep{uren1985}. In scaled CMOS detectors, this trapping process causes
a shift in the relationship between the drain current of the MOS
transistor and the gate voltage, discretely increasing or
decreasing the offset level, which fluctuates as random trapping
and de-trapping of charges, either electrons or holes \citep{martin2020}.

This random noise must be addressed when processing
images taken with these detectors. Unusually low or high pixel
values can cause deviations in photometric measurements of
sources in low light levels. Simply averaging images is not an
optimal method because it does not avoid, but may reduce,
RTN. For instance, the pixel “Salt” in the Figure \ref{fig:rtn} shows
fluctuations only toward higher, not lower, values, and
therefore any averaging that is done by including any of the
anomalous points (25.5\%) will be inaccurate. The way to deal
with these outliers is to take several images and stack them
using a more robust statistic. This technique will be discussed
further in Section \ref{sec:discussion}. It is also possible to identify some pixels
with higher than average signals but with low standard
deviation. Some of them are independent of exposure time,
such as the pixel “High” in the figure, which can be corrected
by subtracting the master bias. However, others depend on
exposure time and are treated as warm pixels in the next
section.

\subsection{Dark current and warm pixels}\label{sec:dark}
Dark current (DC) refers to the unwanted leakage current
that is generated in photosensitive devices in the absence of
incoming light, which is mainly due to the thermal generation
of charges in the silicon layer, and is strongly dependent on the
temperature and exposure time. To characterize the DC, five dark frames of 1000 s exposure time were taken, a master bias
taken just before was subtracted, and they were median
stacked. This was done for various temperatures: from $5\degree$C to $-10\degree$C on the QHY600M Pro and from $-5\degree$C to $-20\degree$C on the
QHY411M, on which water cooling was installed for these
tests. The DC distribution in electrons per pixel and seconds of
exposure is shown in Figure \ref{fig:dark_hist}. The distribution curves have
similar shapes, with a shift toward higher DC with increasing
temperature. When the peak is reached, the DC drops rapidly
toward a smooth hump, from which it drops back down again.
This hump corresponds to a set of pixels with exceptionally
higher DC, which are seen in the images as pixels with a higher
signal that is steady from frame to frame, unlike the Salt \&
Pepper effect described in the previous section, which increases
with exposure time. At $-10\degree$C, only 0.024\% of the QHY600M Pro pixels and 0.005\% of the QHY411M have a DC greater
than RON. These warm pixels have signals lower than the
saturation level, and therefore they can be corrected with an
appropriate master dark. The median value of the dark current
for each camera and temperature is shown in Table \ref{tab:dark}. It is
worth mentioning here that no glow has been seen in any of the
darks taken in either of the cameras, as reported for other
sCMOS sensors \citep{Karpov2021}.\\

\begin{figure}
  \centering
    \includegraphics[width=\linewidth]{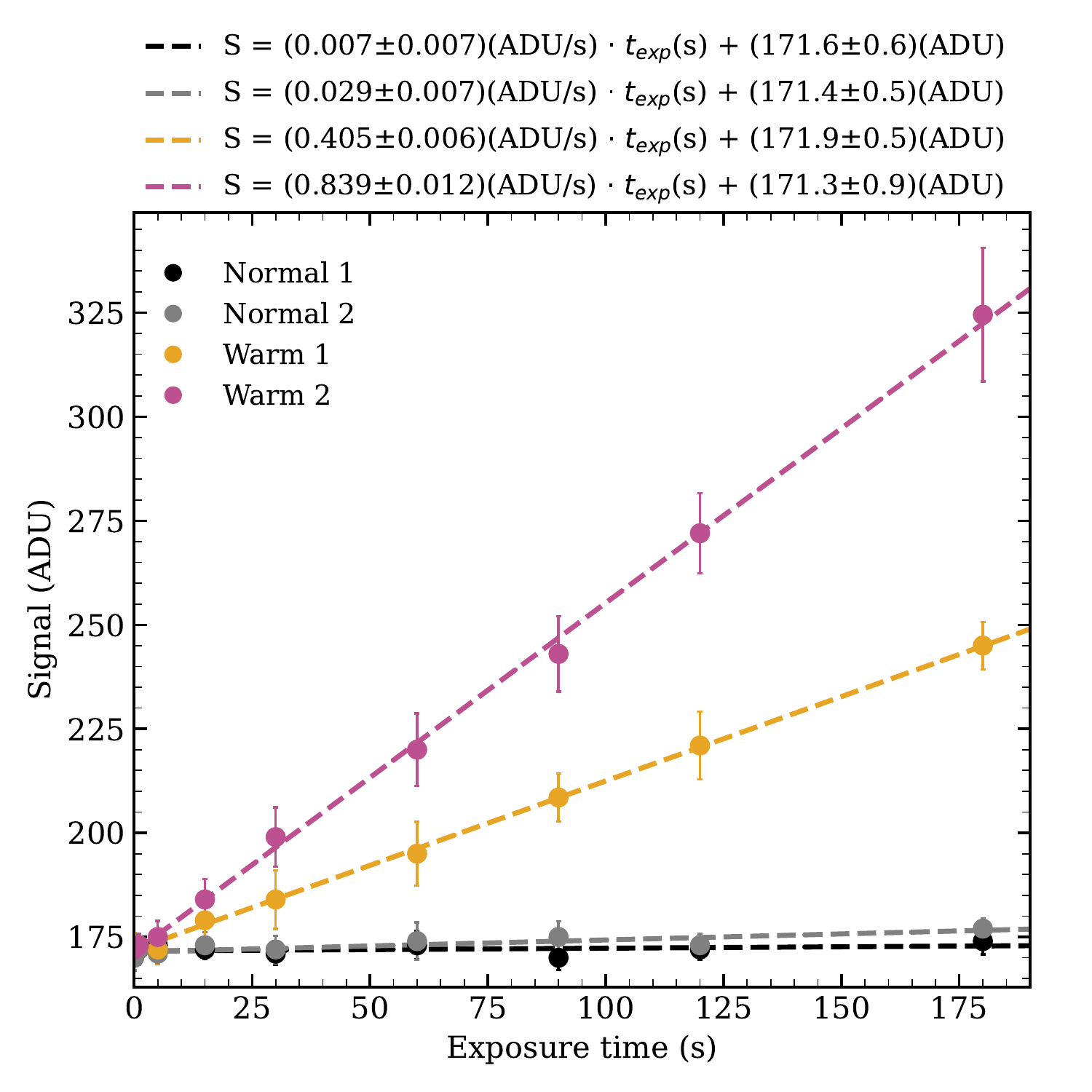}
      \caption{Signal of two warm and two normal pixels of the QHY411M as a
function of exposure time, obtained as the median value of these pixels over
sets of 21 frames taken consecutively, with the standard deviation included as
its uncertainty. The parameters of the least squares fitted lines are shown at
the top.}\label{fig:warm_expt}
\end{figure}

\begin{figure}
  \centering
    \includegraphics[width=\linewidth]{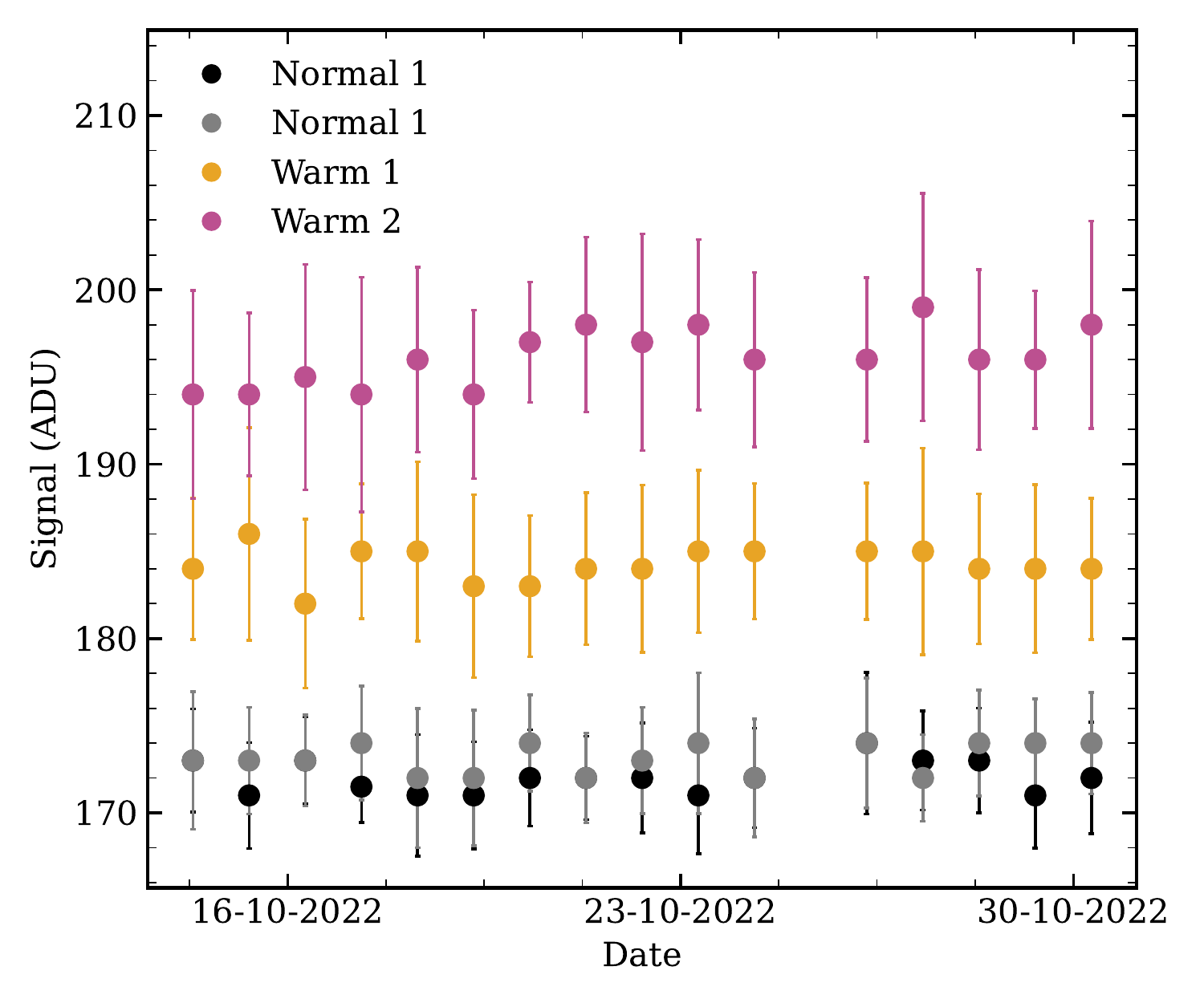}
      \caption{Signal of two warm and two normal pixels of the QHY411M
obtained daily for more than three weeks, obtained as the median of the value
of these pixels in sets of 21 frames of 30 s exposure taken consecutively,
including the standard deviation as its uncertainty.}\label{fig:warm_time}
\end{figure}

\begin{deluxetable}{ccc}
\tablecaption{Median dark current (DC) for different cooling temperatures for each camera.}\label{tab:dark} 
\tablehead{
\colhead{T ($\degree$C)} & \multicolumn{2}{c}{DC (e$^-$ px$^{-1}$ s$^{-1}$)}\\
 & \colhead{QHY600M Pro} & \colhead{QHY411M}}
\startdata 
5   & 0.025 & \\
0   & 0.018 & \\
-5  & 0.014 & 0.009\\
-10 & 0.011 & 0.007\\
-15 &       & 0.005\\
-20 &       & 0.004\\
\enddata
\end{deluxetable}

Some warm pixels in a central region of the QHY411M have
been identified and revised in more detail. First, sets of 21 dark
frames taken consecutively with different exposure times have
been stacked. The median signal of two warm and two normal
pixels is shown in Figure \ref{fig:warm_expt}, where the standard deviation is
included as uncertainty bars. It is clear that the signal of these
warm pixels scales linearly with the exposure time in the range
of 1--300 s. The result of the linear least squares fit is shown at
the top of the plot, all of them with $R^2>0.998$. Note that in this
case a master bias has not been subtracted, and therefore the
value of the offset is different from 0. This result implies that it
is possible to correct the warm pixels in images of a given
exposure time if the slope of this line for each pixel is known,
which can be obtained with other exposure times (e.g., at the
beginning of the night). Although a scaled master dark is
intended to be subtracted from the science images, it should be
noted that warm pixels, having a higher thermal signal, will be
noisier than those with a lower DC.

For more than 3 weeks, sets of 21 darks of 30 s exposure
were taken every night. The median value in these four pixels
as a function of time is shown in Figure \ref{fig:warm_time}, where the error bars
again show their standard deviation. With this test, we have
tried to see if there is a substantial variation in the values of
these pixels over longer periods of time. Taking into account
that the noise in these pixels is a combination of the RON and
the thermally generated electrons, which follow a Poissonian
distribution, it can be seen that the possible variations in the pixel dark signal over at least these three weeks is not
significant with respect to the total noise. Therefore, meaningful
night-to-night variations of the master dark or bias are
not expected.

\begin{figure*}
    \centering
        \includegraphics[width= 0.49\linewidth]{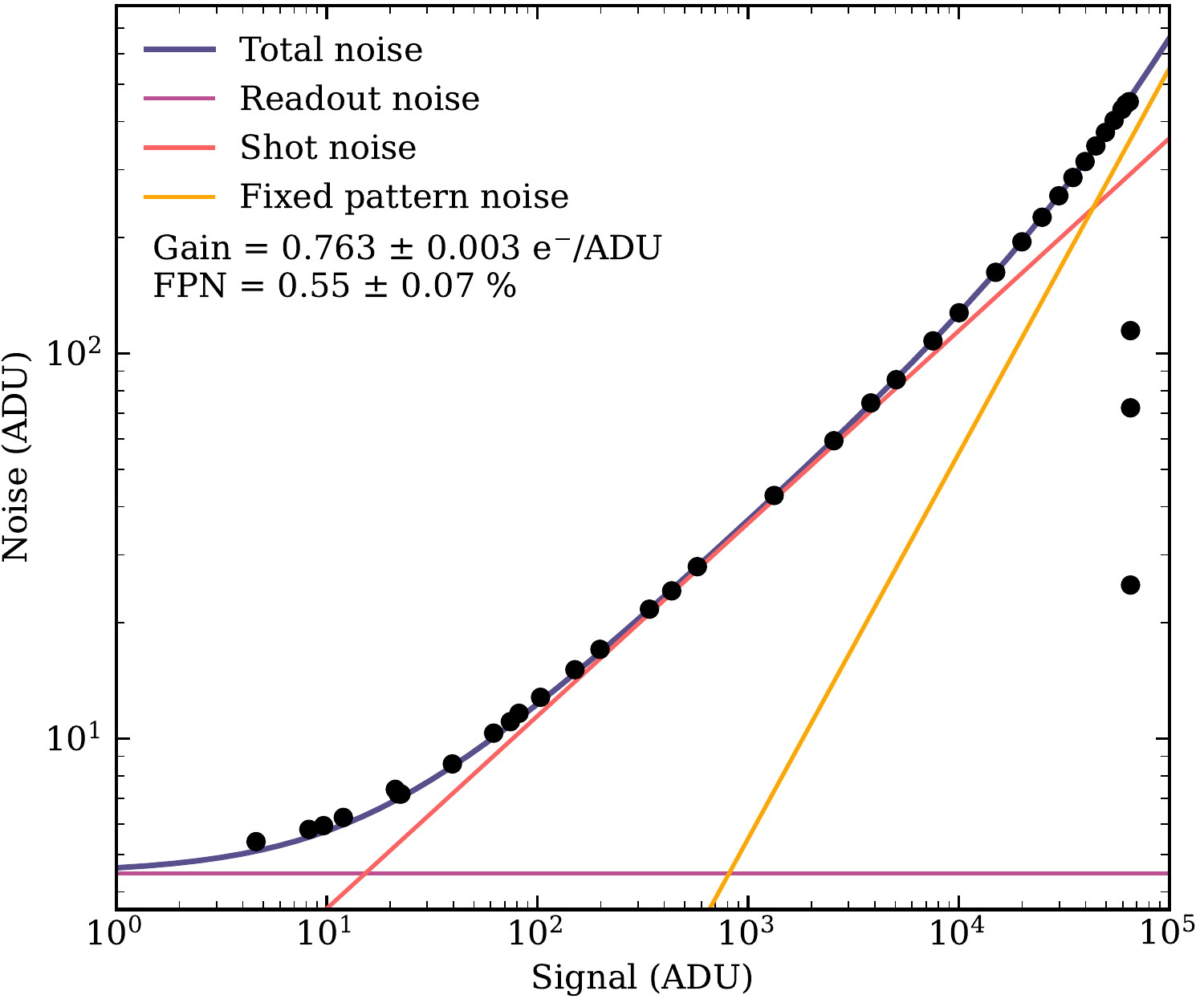}
\includegraphics[width= 0.49\linewidth]{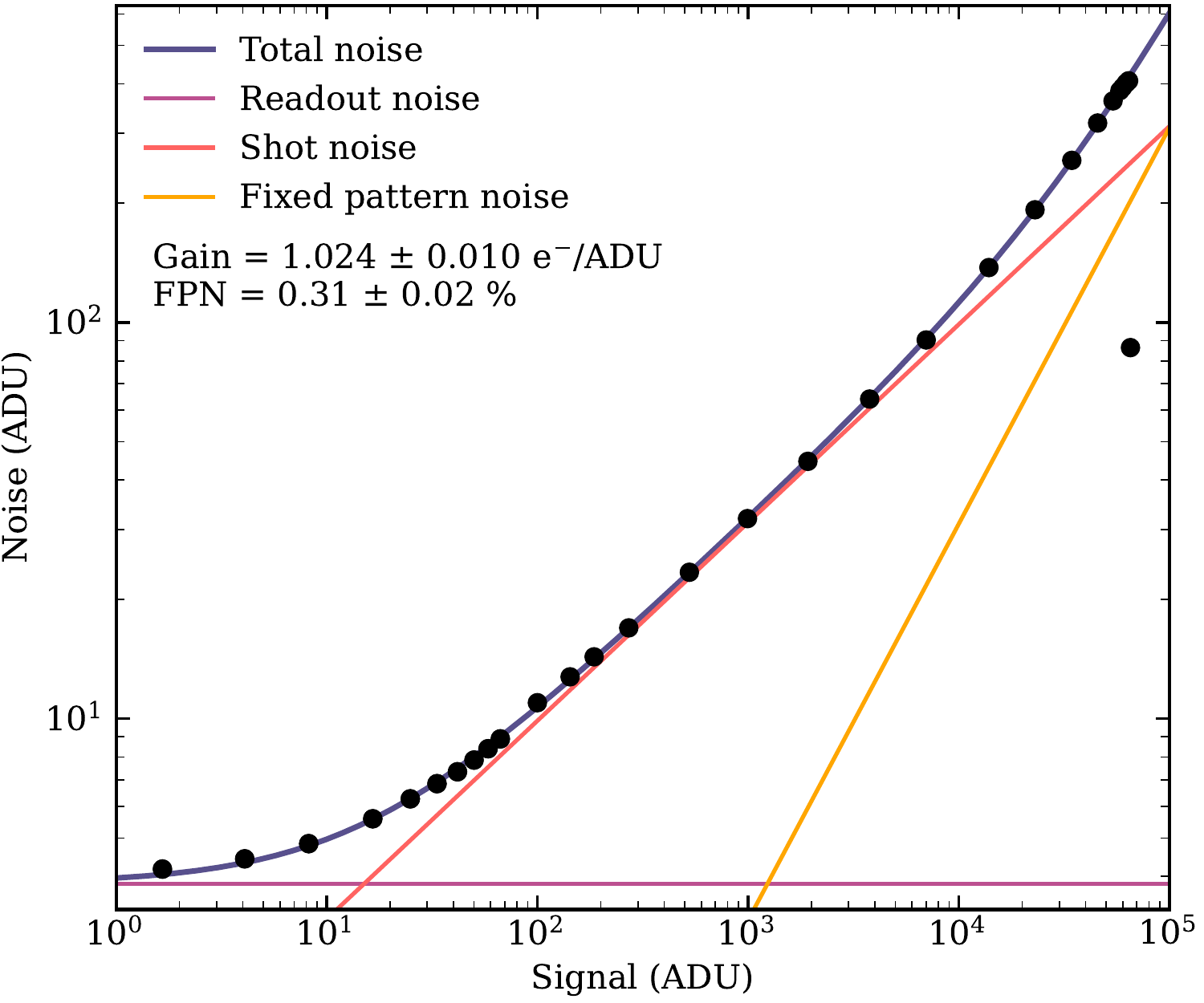}
\caption{Photon transfer curves obtained from sets of three bias-subtracted images, median stacked, and taken under uniform illumination with increasing exposure
times. The QHY600M Pro Mode \#1@0 (left-hand panel) and the QHY411M Mode \#4@0 (right-hand panel). The principal components of the noise have been
distinguished as colored lines, whose characteristic values are also included. The gain and FPN has been obtained from the fitting to the total noise function.}\label{fig:PTC}
\end{figure*}

\begin{figure*}
    \centering
\includegraphics[width= 0.49\linewidth]{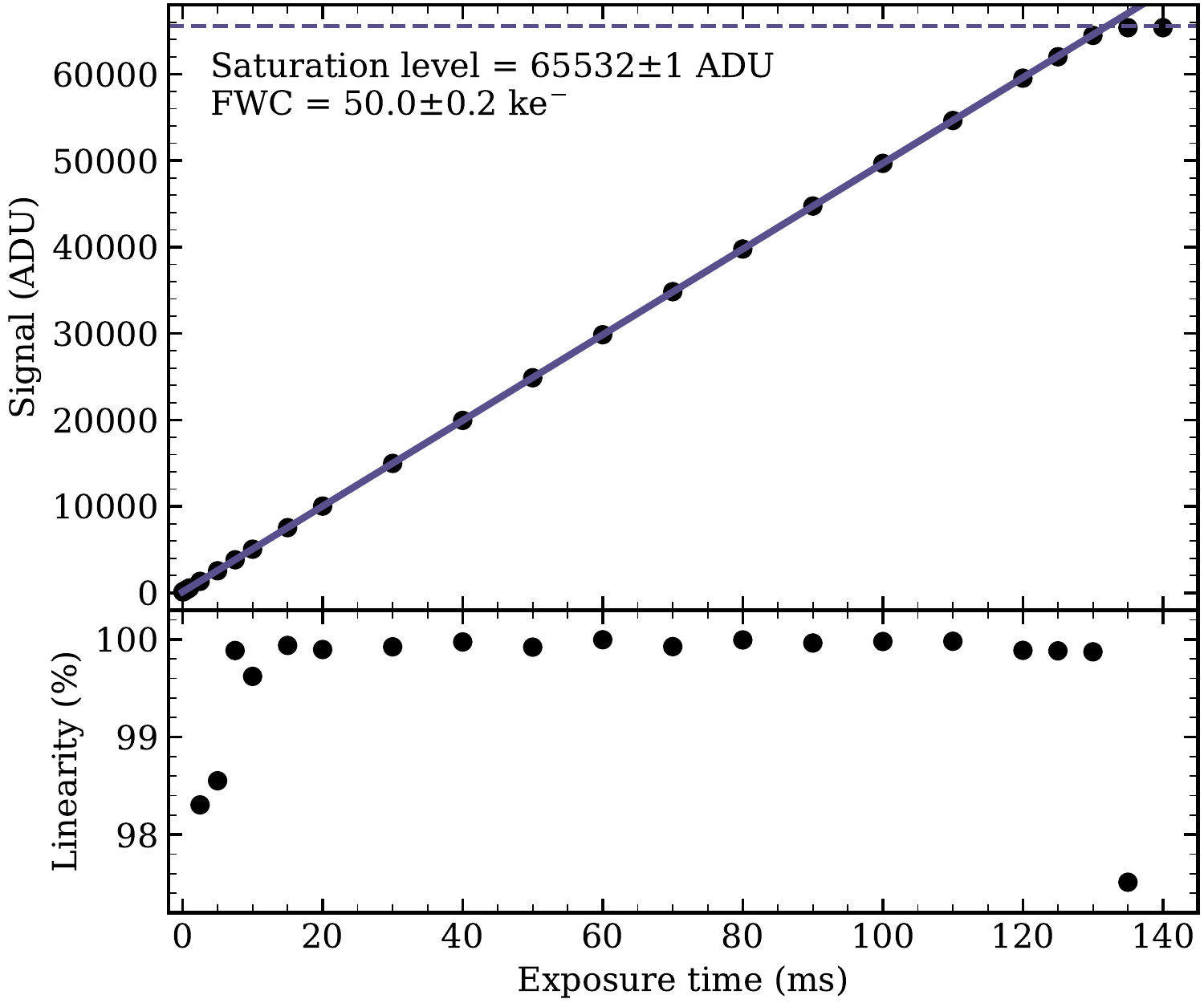}
\includegraphics[width= 0.49\linewidth]{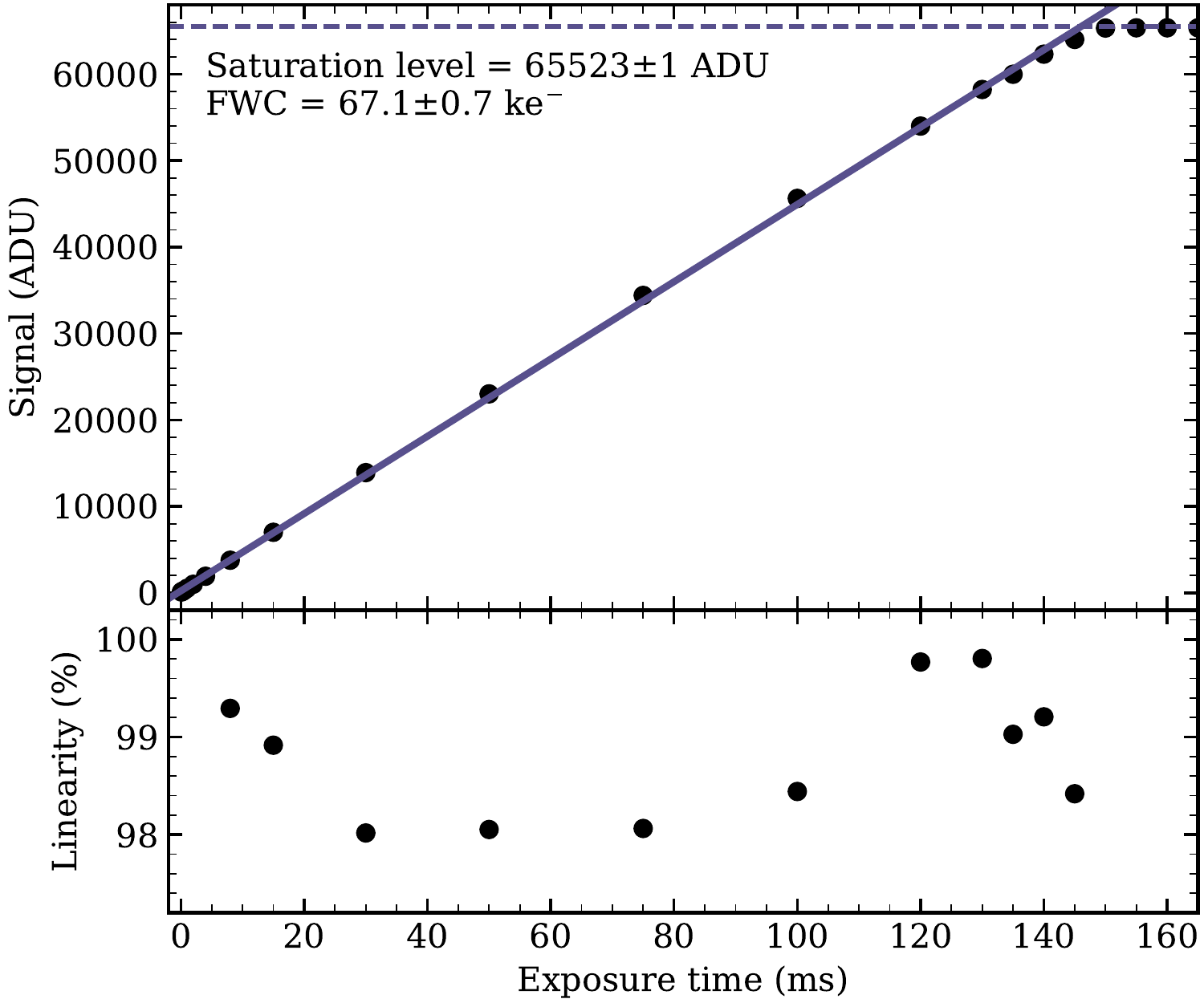}
\caption{Average signal in sets of three images taken with increasing exposure time (top) for the QHY600M Pro Mode \#1@0 (left-hand panel) and the QHY411M
Mode \#4@0 (right-hand panel). The saturation level is indicated by a horizontal-dashed line. A straight line was fitted to points with values between 100 and 60,000
ADU. The relative rms of the differences between the measurements and the fitted line is shown in the lower plots.}\label{fig:linearity}
\end{figure*}

\subsection{Photon Transfer Curve and linearity}\label{sec:PTC}
The photon transfer curve (PTC) is used to characterize the
response of the sensors to homogeneous illumination, obtaining
main features such as the gain---conversion factor between
electrons and digital counts (ADU), the full-well capacity
(FWC), or the contribution of the different sources to the total
noise. The laboratory set-up that is described in Figure \ref{fig:lab} was
modified by removing the monochromator, so that the light
coming from the QTH lamp could directly enter the integrating
sphere. The power supply was kept stable. Sets of three images
of uniformly illuminated exposures were then taken while
increasing the exposure time until the saturation turn-off point
was reached. Three bias frames were taken before and after
each series, stacked by median 3$\sigma$-clipping and subtracted
from all of the illuminated images. A central region of interest
(ROI) of $4096 \times 4096$ unbiased pixels was used for both
cameras. The signal was obtained as the mean of the three
images stacked. The mean of the standard deviation across the
three frames was taken to obtain the noise value.

The total noise under illumination is given by the quadrature
sum of three main components: RON, shot noise, and fixed
pattern noise (FPN):
\begin{equation}
    \sigma_{\text{TOTAL}} (\text{ADU})= \left[\sigma_{\text{RON}}^2 
+ \frac{S }{G} + \left(P_{\text{FPN}} S \right)^2\right]^{1/2} \label{eq:ptc}
\end{equation}
where $\sigma_{\text{RON}}$ (ADU) is the readout noise, $S$ (ADU) the signal, $G$ (e$^{-}/$ADU) the gain and $P_{\text{FPN}}$ the FPN factor \citep{Janesick2007}.

The relationship between the noise and signal for the two
sensors is shown in Figure \ref{fig:PTC}. The RON was fixed with the
value defined in the previous section, while the gain and FPN
parameters were obtained by fitting the expression (\ref{eq:ptc}) on a
logarithmic scale. This procedure was repeated in windows of $200 \times 200$ pixels along the ROI to obtaining a curve for each of
them, 360 in total. The rms of the parameter distributions was
used to estimate their uncertainty. In both plots, three regions
can be distinguished. At low illumination, below 10 ADU,
RON, which does not depend on the signal, is the dominating
noise source. From then on, shot noise starts to become
important, and is essentially the main component between 100
and 10,000 ADU. Thereafter, the contribution of the FPN
becomes significant and a deviation from Poissonian behavior
is observed. In the QHY600M Pro, this source is the main
contributor above 50,000 ADU, with a factor of $0.55\%\pm0.02\%$ , being lower, $0.31\%\pm0.02\%$, in the
QHY411M, where the shot noise is dominant almost up to
the saturation point. A PTC has also been obtained for other
operating modes, whose resulting values are included in
Table \ref{tab:res}.

The linearity of the sensors is evaluated by plotting the
signal (adding the bias level) against the exposure time, as is
done for the standard operating modes in Figure \ref{fig:linearity}. A straight line was fitted to the points between 100 and 60,000 ADU (top
plot). The relative rms difference between the points and the
line is shown below. In both cases, the deviation from linearity
is less than 2\% up to the saturation point, which was
determined for all of the operating modes and included in
Table \ref{tab:res}.

\begin{table}
\centering
\caption{Main features of the different operating modes of the QHY600M Pro (top) and QHY411M (bottom), obtained in laboratory tests. including the gain, readout noise (RON), fixed pattern noise (FPN) and full-well capacity (FWC).}\label{tab:res} 
\renewcommand{\arraystretch}{1}
\begin{tabular}{ccccc}
\hline\hline
Mode & Gain & RON & FPN & FWC \\
 & (e$^-$/ADU) & (ADU) & (\%) & (ke$^-$)\\
\hline
\multicolumn{5}{c}{QHY600M Pro}\\
\hline
0@26 & 0.405$\pm$0.003 & 6.28 & 0.521$\pm$0.005 & 25.3$\pm$0.2\\
1@0 & 0.763$\pm$0.003 & 4.56 & 0.55$\pm$0.07 & 50.0$\pm$0.2\\
\hline
\multicolumn{5}{c}{QHY411M}\\
\hline
1@0 & 0.979$\pm$0.008 & 3.03 & 0.29$\pm$0.02 & 27.2$\pm$0.2\\
4@0 & 1.024$\pm$0.010 & 3.74 & 0.31$\pm$0.02 & 67.1$\pm$0.7\\
5@0 & 0.332$\pm$0.002 & 4.65 & 0.32$\pm$0.02 & 22.8$\pm$0.1\\
6@0 & 0.776$\pm$0.003 & 4.48 & 0.32$\pm$0.02 & 50.6$\pm$0.2\\
7@0 & 0.251$\pm$0.001 & 5.69 & 0.34$\pm$0.02 & 16.50$\pm$0.05\\
\hline\hline
\end{tabular}
\end{table}

\subsection{Quantum efficiency}\label{sec:QE}
Using the optical setting described in Figure \ref{fig:lab}, central
wavelengths in the range between 350 and 1100 nm with 25 nm
steps were selected in the monochromator. The grating
configuration was set to have an outcoming light with $\pm1$ nm bandwidth. The QHY600M Pro was placed at 13.3 mm from
the exit of the dark box, which is separated by 30 mm from the
Hamamatsu S2281 photodiode. The total distance between the
IMX455 sensor and the photodiode, considering the back focus
distance of the camera, was 66.8 mm. In the case of the
QHY411M, it could be placed in contact with the box, so the
total distance between the photodiode and the IMX411 sensor
was 58.5 mm. Both cameras were binned $4 \times 4$ to have enough
signal with exposure times shorter than 10 s at those
wavelengths where they are less efficient.

Three images were taken at each wavelength step, a master
bias that was created at the beginning of the series was
subtracted from all of them and they were stacked with the 3$\sigma$-clipped median. Simultaneously, the output intensity of the
photodiode placed at the secondary port of the integrating
sphere was measured with the picoammeter for approximately
30 s, taking an average value. The observed fluctuations in this
value were always less than 2\%. The systematic uncertainty of
the method is estimated to be around 2\%. To avoid vignetting
effects at the edges of the dark box-camera junction, a central
ROI of $1000 \times 1000$ binned pixels was used to obtain the
median signal $S$, after checking that there were no inhomogeneities
in the sensor illumination in that area. The exposure
time $t_{\text{exp}}$ had to be varied throughout wavelength steps to keep
all of the measurements in the shot noise dominated region between 1000 and 40,000 ADUs. The absolute quantum
efficiency for each wavelength is given by the expression:
\begin{equation}
    \text{QE}(\%) = \frac{S\text{(ADU)}\cdot G(\text{e}^{-}/\text{ADU})}{t_{\text{exp}}\text{(s)}\cdot F (\text{e}^{-}/\text{s})}\cdot 100
\end{equation}

\begin{figure}
    \centering
    \includegraphics[width=\linewidth]{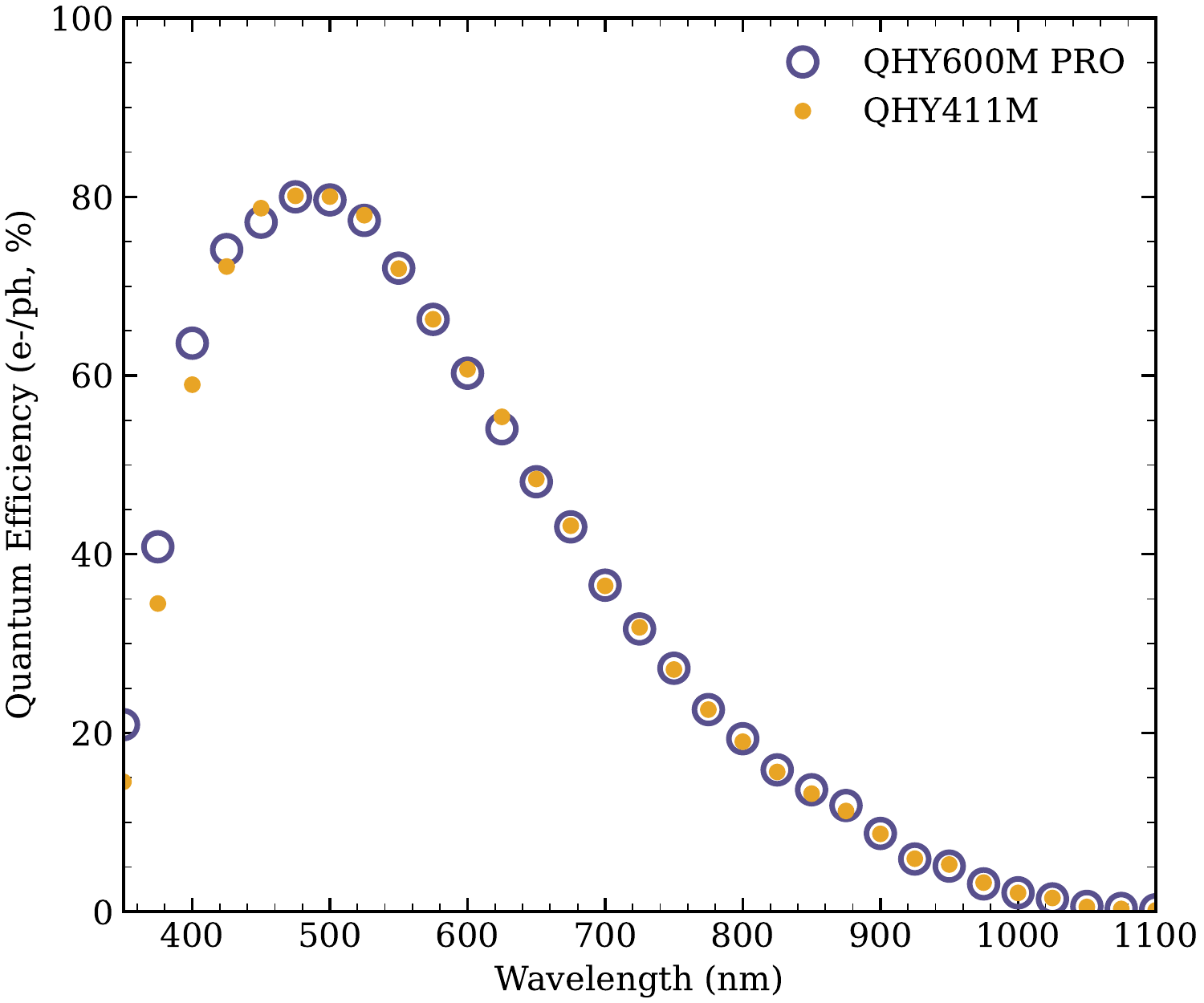}
    \caption{Absolute quantum efficiency curves of the QHY600M Pro (blue-open circles) and the QHY411M (yellow dots). The systematic uncertainty is estimated to
be around 2\%}
    \label{fig:QE}
\end{figure}

\begin{figure*}
    \centering
    \includegraphics[width=\textwidth]{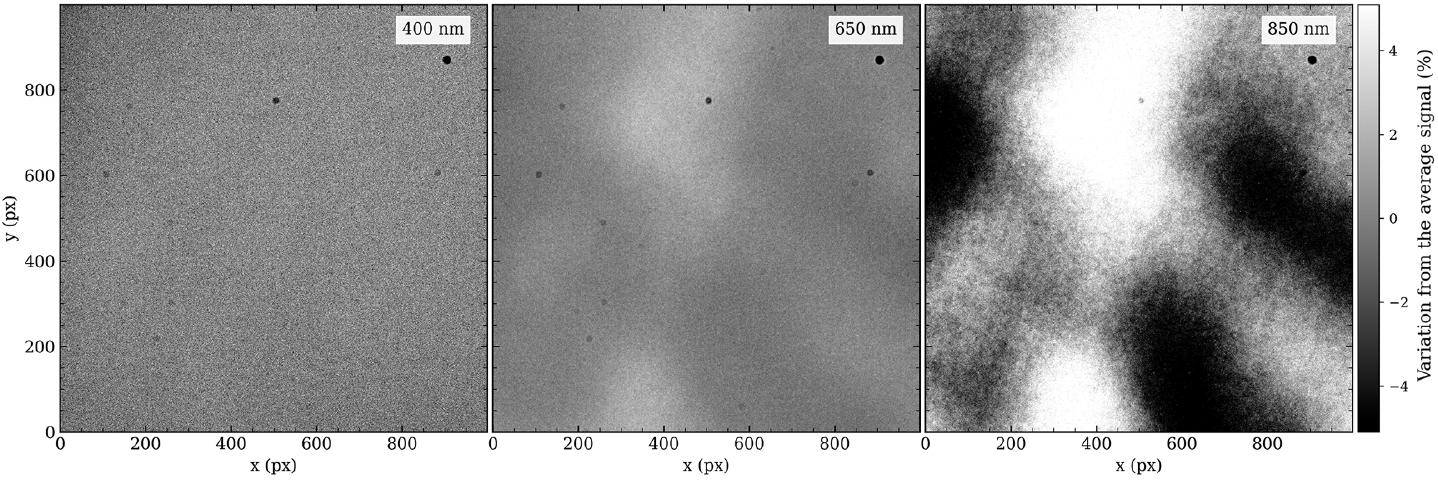}
    \caption{Etaloning pattern observed when illuminating the QHY600M Pro in
a central ROI of $1000 \times 1000$ pixels $4 \times 4$ binned with $\pm1$ nm bandwidth light at different wavelengths. Similar behaviour is observed in the QHY411M.}
    \label{fig:etaloning_qhy600}
\end{figure*}

The QE curves for the sensors are shown in Figure \ref{fig:QE}. Both
show very similar behavior, with a peak efficiency of 80\% at
475 nm, a steep drop at shorter wavelengths, and a gradual
decrease toward the redder ones, with a QE of 40\% at 700 nm
and 10\% at 900 nm. Back-illuminated sCMOS sensors with
reduced pixel size, such as the IMX455 and IMX411, have a
typical silicon substrate thickness of around 3 $\mu$m \citep{Yokogawa2017}. This optimizes the photon absorption in the visible
range but makes the less energetic photons, which have a
higher penetration capability, less likely to be detected.
Without any additional red enhancement technology, improving
efficiency in the red and near-infrared requires a thicker
substrate that, with such small pixels, would lead to image
degradation owing to crosstalk between adjacent pixels. This is
the reason for the poor performance at longer wavelengths.

It should be noted that the curves that are obtained here
represent an overall reduction of 9\% over the full absolute QE curve published by QHYCCD, which has a peak of 92\% at
450 nm, 46\% at 700 nm, and 15\% at 900 nm. Nonetheless, the
conditions and configuration used by them, as well as
the uncertainties in their measurements, are unknown and the
calibration method to obtain the QE curves is different from
the one used here, which has been used to calibrate multiple
astronomical instruments before. The same procedure and test
bench have been used, for instance, to calculate the QE in
sCMOS cameras such as Andor Marana, FLI-Kepler, and
ORCA-Hamamatsu. They have also been used in several deep
depletion CCDs, such as the well-known Teledyne e2v 4482
and 231-84 BI. In all cases, QE fits rather well with data
supplied by the manufacturers. \cite{Gill2022} have very
recently presented a low-cost method to obtain, among other
features, the absolute QE of detectors being applied to the
IMX455 sensor. Their results also show a lower performance
than that presented by QHYCCD, especially in the red part,
although they had a similar peak efficiency, with 93\% at
480 nm, 42\% at 700 nm, and 10\% at 900 nm, making an overall
deviation from our results of 5\%. Finally, \cite{Betoule2022} have studied the quantum efficiency of the IMX411 sensor in
great detail and obtained a result that is very similar to Figure \ref{fig:QE}, with a peak of 83\% at 490 nm, 37\% at 700 nm and 6\% at 900 nm.

During the tests, optical etaloning was observed at longer
wavelengths, as shown in Figure \ref{fig:etaloning_qhy600} for the QHY600M Pro.
Similar behavior was observed also for the QHY411M and
reported by \cite{Betoule2022}. The maximum variation over
the average frame mean value is around 1\% at 650 nm, and
reaches 10\% at 850 nm and above. This is a known effect in
thinned back-illuminated devices, resulting from multiple
reflections produced inside the depletion region by a mismatch
between its refractive indices and the adjacent layers. This
effect has not been observed in sky images taken with TAR04
using a wider bandwidth, such as SDSS i$^\prime$ (690--850 nm).

\subsection{Charge persistence}\label{sec:persistence}
Charge persistence is an effect that occurs when a portion of
the signal remains in the detector element after the sensor has
been read out. It is a consequence of the creation of traps at the
interface between the photodiode and the transfer gates, which
capture free electrons and gradually release them, resulting in a
decay of the residual signal in the subsequent images after the
illuminating source has been removed. This effect has been
observed with tests performed in the laboratory with the FLI
Kepler KL400 and Andor Marana cameras, both with
GSENSE400BSI sensors, showing a behavior similar to that
reported by \cite{Karpov2021}. Although they did not
observe any effect in sky images, previous tests performed with
these cameras on the same telescopes used in our work did
show a smearing effect in pixels that had previously been
exposed close to the FWC, which remained visible up to tens of
minutes later.

\begin{figure}
    \centering
    \includegraphics[width=\linewidth]{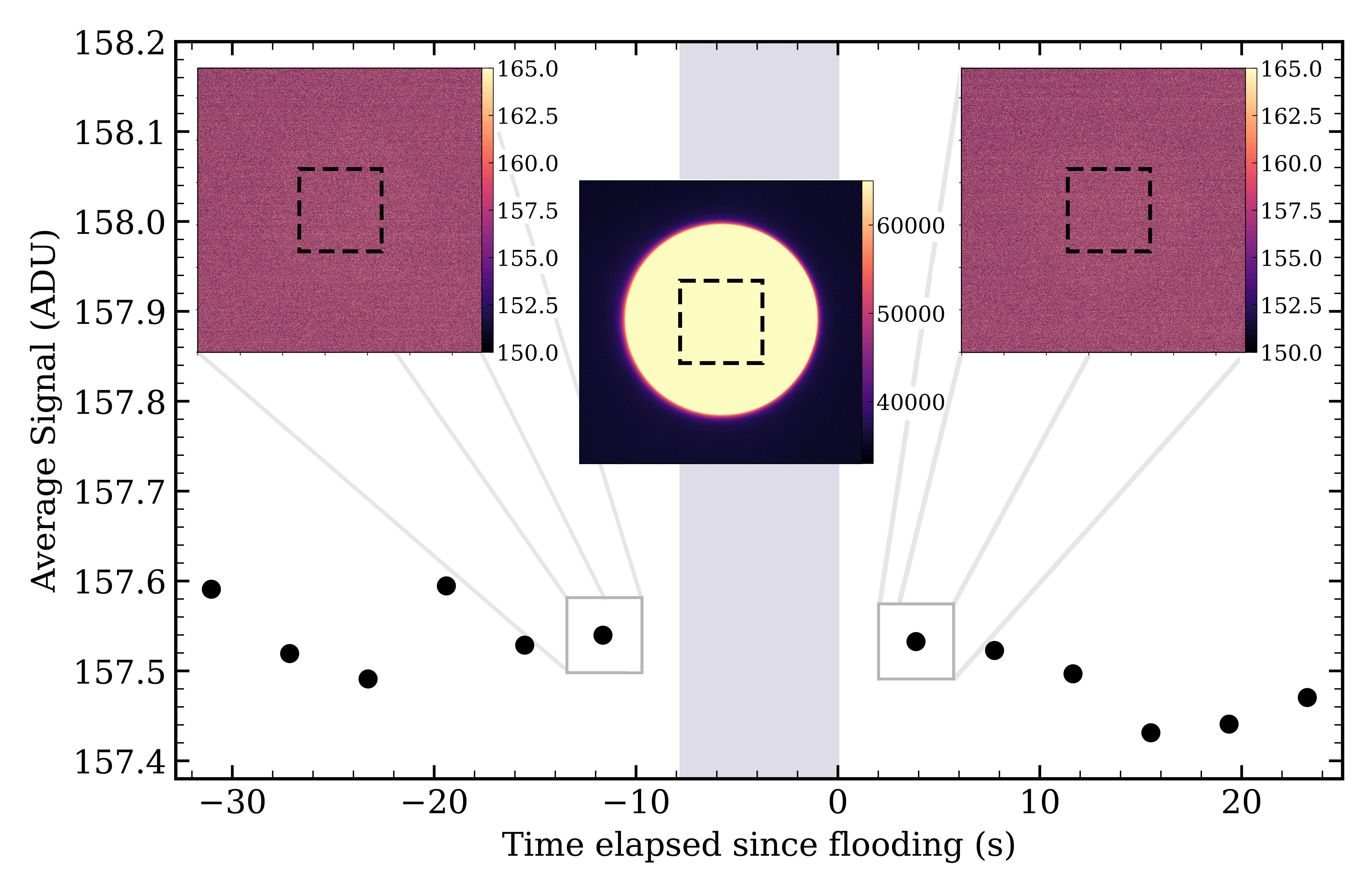}
    \caption{Average signal (black dots) in 1 second frames taken continuously
during sensor flooding (central shaded area). Images before (left-hand), during
(center), and after (right-hand) are shown at the top. The region in which the
average signal has been calculated is indicated by dashed rectangles.}
    \label{fig:persistence}
\end{figure}

This effect has been tested in the laboratory for both sensors.
To do so, following the same set-up used in Section \S\ref{sec:PTC}, i.e., removing the monochromator and exposing the cameras to the
QTH lamp light, a pinhole was placed inside the black box, just
in front of the sensor. Consecutive 1 s images were taken with
the shutter closed. The shutter was then opened for about 10 s,
saturating the illuminated pixels, and closed again. This
sequence is shown in Figure \ref{fig:persistence}. On the top are the images
taken with the QHY600M Pro before the shutter was opened
(left-hand), during the exposure to the light coming from the
pinhole (center), and just after the shutter was closed again
(right-hand). In the central region of the illuminated spot area,
indicated by a dashed square, the mean signal has been
measured and is shown in black dots below. The signal before
and after the flooding shows a stable trend, with very small
oscillations, not a sharp increase in the signal right after the
exposition followed by a slow decay to the original level, as
shown in Figure 10 of \cite{Karpov2021}. Hence, we
conclude that the IMX455M and IMX411M sensors do not
exhibit charge persistence.

\section{On-sky tests}\label{sec:sky}
\begin{figure}
    \centering
    \includegraphics[width=\linewidth]{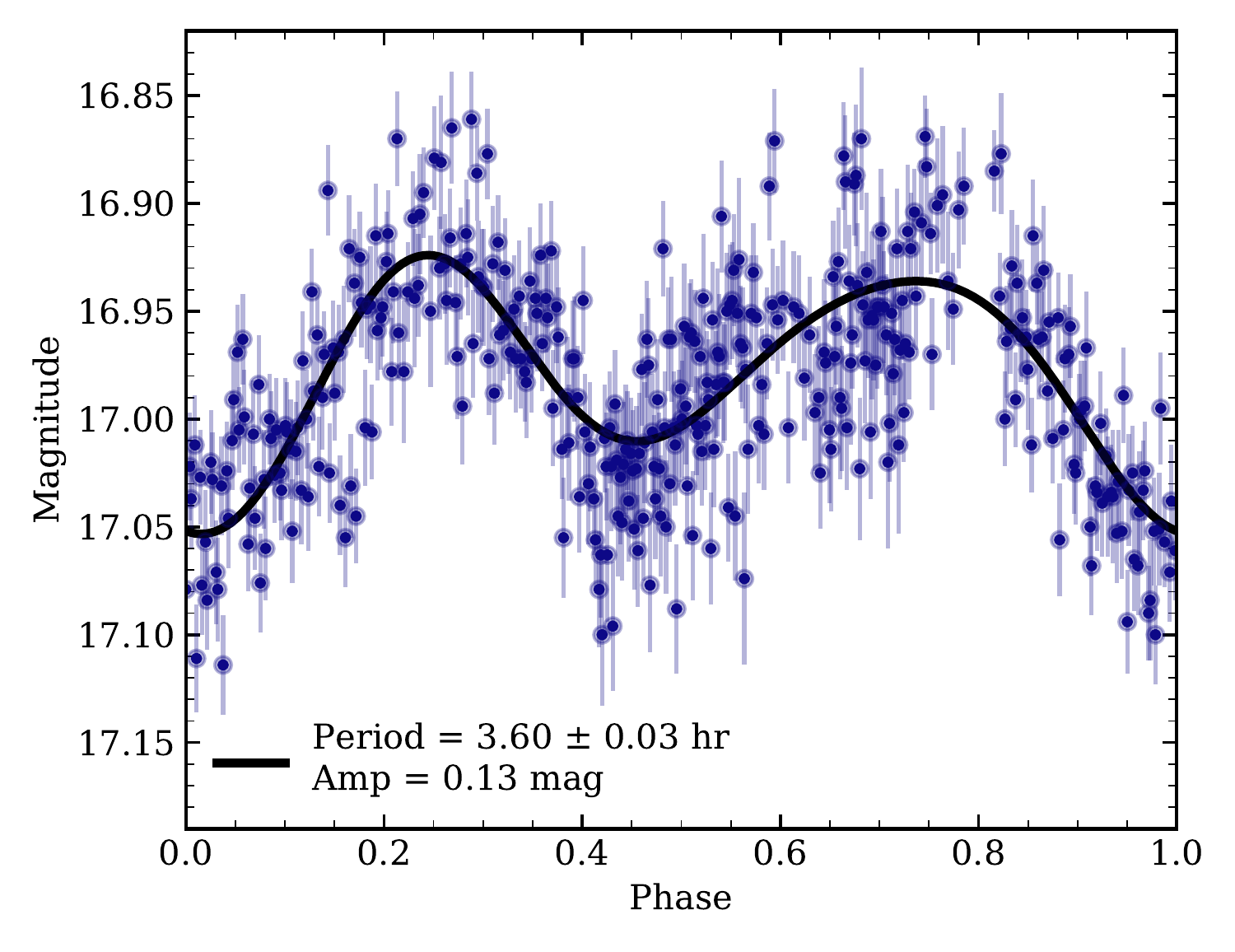}
    \caption{Phased light curve of the asteroid (3200) Phaethon, obtained with
the QHY600M mounted at the prime focus of a 0.46-m $f$/2.2 telescope, no filter
installed.}
    \label{fig:phaethon}
\end{figure}

\begin{figure}
    \centering
    \includegraphics[width=\linewidth]{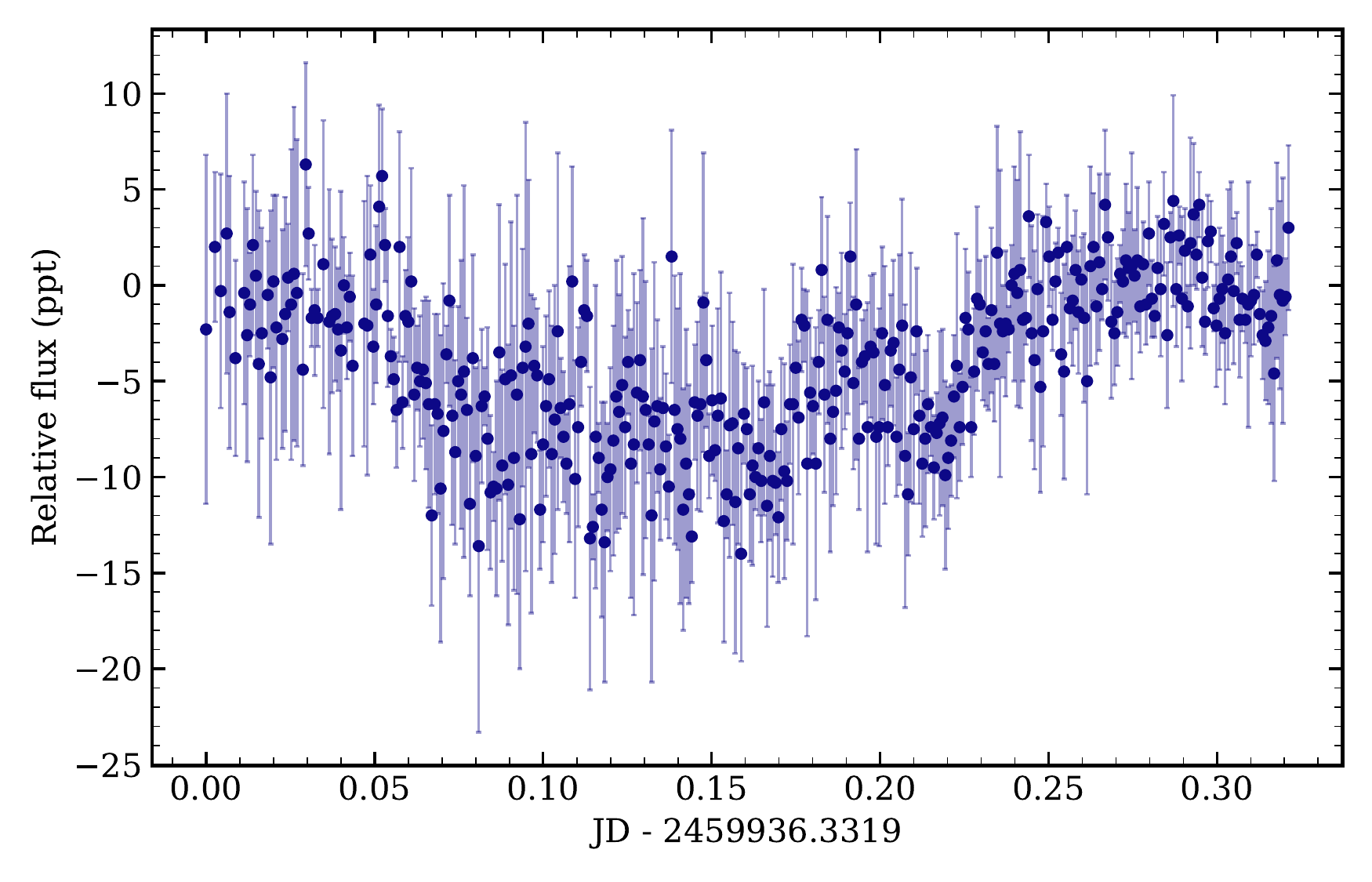}
    \caption{Transit of the exoplanet TOI-1135 observed with QHY411M
mounted on a Nasmyth focus of the TTT-2 telescope at 0.80-m and $f$/6.85,
SDSS g$^\prime$ filter.}
    \label{fig:exoplanet}
\end{figure}

Both cameras have been extensively tested on images taken
with telescopes, obtaining photometric accuracies as expected
for the characteristics described above. The QHY600M Pro
was installed at the prime focus of a 0.46-m $f$/2.2 telescope. Figure \ref{fig:phaethon} shows the phased light curve of asteroid (3200)
Phaethon, which was observed for 6 consecutive hours. At the time of observation, the object had an apparent magnitude of
$V=17$ and was moving at a speed of 1$^{\prime\prime}$.6 minute${-1}$. The
photometric uncertainties that we obtained are in the order of
hundredths of a magnitude and the rotation of the asteroid, with
an amplitude of 0.13 mag, is clearly detectable. This camera is
a very suitable choice for very fast telescopes with primary
focus because its small size and compact shape drastically
reduce obscuration. With a pixel size of 3.76 $\mu$m, it also allows
us to obtain a plate scale that is very suitable for sites with
excellent seeing, such as the Teide Observatory. Furthermore,
given its low readout noise and negligible readout time, it is
possible to take continuous short frames and combine them by
aligning with the object or the stars, thus allowing fainter
objects to be reached with very little time lost. This also allows
the study of, for instance, very fast rotating objects with good
temporal sampling \citep{Licandro2023b} or using “shift-and-add”
techniques, such as synthetic tracking \citep{shao2014}, to improve the detection performance of faint fast-moving
objects.\\

Figure \ref{fig:exoplanet} shows a transit of the exoplanet TOI-1135 that was
observed with the QHY411M mounted on one of the Nasmyth
foci of the TTT-2 telescope, 0.80-m $f$/6.85, SDSS g$^\prime$ filter, with
a plate scale of 0$^{\prime\prime}$.14 px$^{-1}$. Groups of four images of 5 s
exposure were stacked to improve the SNR. The standard
deviation of the differential aperture photometry measurements
in pre-transit, during, and post-transit is 2.2, 3.3, and 2.0 mmag,
respectively (M. Mallorquin et al., in preparation). With such a
small plate scale, the psf was oversampled, with approximately
8 px of FWHM, spreading the flux of the star over a larger
number of pixels. For very bright targets, such as this $V=9.6$ mag star, this allows slightly longer exposures to be taken
without reaching the saturation point, thus reducing the random
and flat-fielding errors of the telescope that, with a larger scale,
would perhaps need some defocusing \citep{Southworth2009}. In addition, by taking short exposures, the time lost on a CCD
could be equivalent to, or even longer than, the exposure time,
which makes it very inefficient. With the QHY411M, the
exposure time could be extended, thus improving photometric
accuracy, and the readout time is almost non-existent. Although
such small plate scales are generally undesirable, they allow
these cameras to be used in other scientific applications (e.g.,
fast-moving object astrometry or lucky imaging).

\begin{figure*}
  \centering
    \includegraphics[width=.95\linewidth]{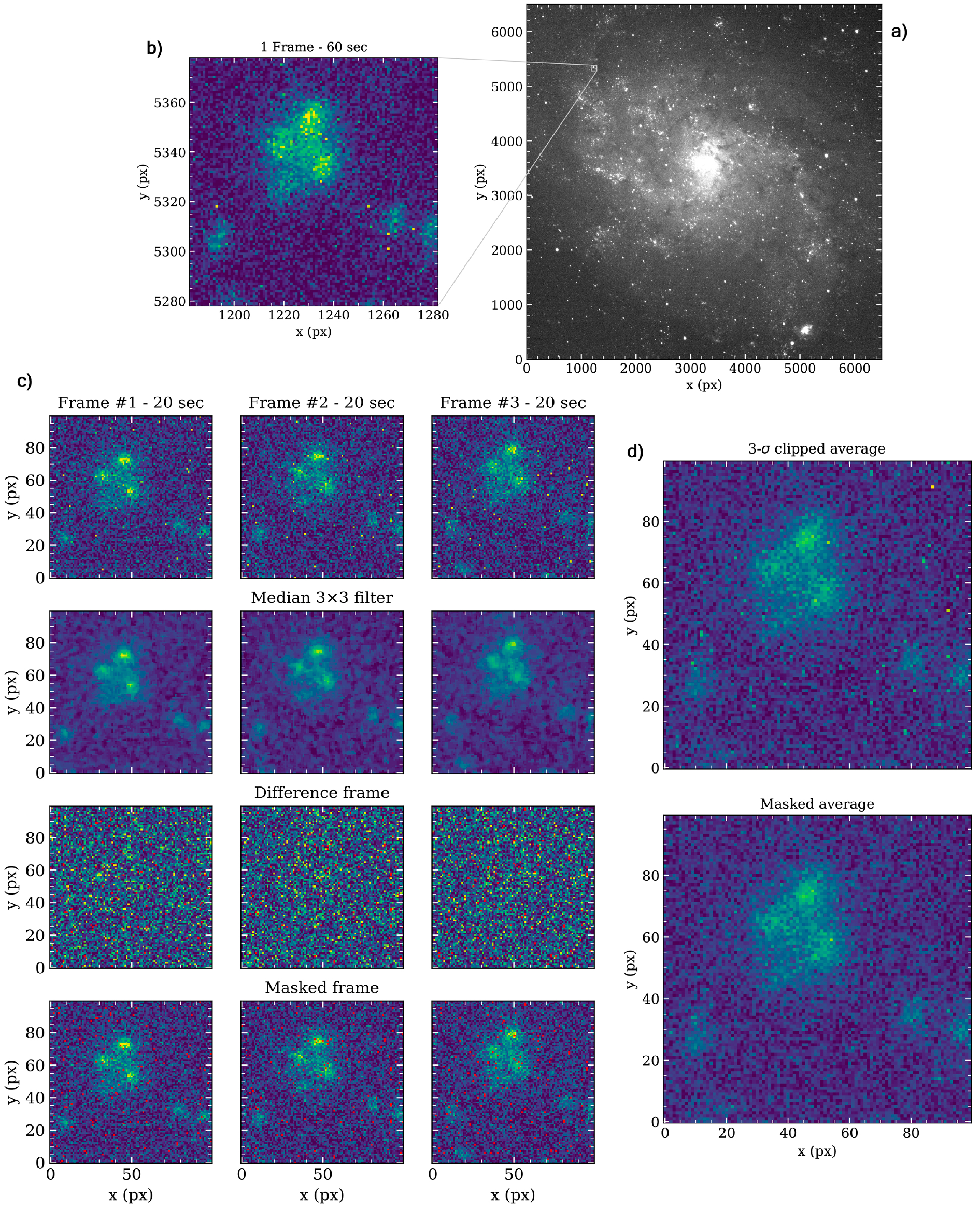}
      \caption{(a) Image of M33 taken with the QHY411M at the TAR04 telescope with an exposure time of 60 s. (b) Zoomed region $100 \times 100$ pixels where the Salt \& Pepper effect is visible. (c) From top to bottom: $3 \times 20$ s sequence of the same zoomed region; convolution with a $3 \times 3$ median filter; difference between the
frames in the previous rows, showing the S\&P contaminated pixels in red; and raw frames with those pixels masked in red. (d) Result of stacking the three frames with
an average 3$\sigma$-sigma clipping (top) and with an average after masking the S\&P affected pixels.}\label{fig:sky}
\end{figure*}

\section{Discussion}\label{sec:discussion}
The two instruments with sCMOS sensors that are analyzed
here present characteristics that are compatible with their use in astronomy: they are linear over the whole dynamic range, have
a high full-well capacity, and are slightly affected by dark
current, despite being able to work at higher temperatures than
CCDs. Regarding the quantum efficiency, although the curve
obtained here is slightly lower than that reported by the
manufacturer, 80\% at 500 nm is an acceptable performance for
many scientific programmes and is in general similar or better
than other CCD sensors in the same cost range. An
improvement in efficiency toward redder wavelengths should
be achieved in the next few years, so that sCMOS sensors can
be used on a wider variety of observational targets.

The small pixel size means that these sensors are generally
not the best solution for slow focal length systems, except for
dedicated programmes such as high spatial resolution or lucky
imaging. Binning in sCMOS sensors is done after exposure, so
it does not improve the readout noise or frame rate. In general,
in case it is needed, it is better to do this by software after the
exposure, so that the statistics can be preserved and the 16 bit
limit is not reached. Nevertheless, they can be very valuable in
fast telescopes with larger fields and higher plate scales, which
allows better sampling of the PSF. In addition, their manageability,
and small size and weight are very interesting, e.g., for
prime focus telescopes.

The S\&P effect is one of the main issues that affect the use
of these IMX455 and IMX411 sensors for photometric
measurements. Figure \ref{fig:sky}(a) shows an image of the galaxy
M33 taken with the QHY411M on the TAR04 telescope, with a
UV/IR-Cut/L filter, in a single exposure of 60 s. When
zooming in on a small area of 100$\times$100 pixels (b), several
pixels are observed with values that are clearly higher than
those of their neighbors, as described in Section \ref{sec:bias}. This effect
could influence the photometry of faint objects because the
amplitude of this random fluctuation may be comparable to the
source signal in the pixel.

SSeveral strategies can be pursued to mitigate this problem,
especially by using several consecutive frames. Figure \ref{fig:sky}(c) shows a sequence of $3 \times 20$s, where the S\&P is clearly visible
in each of them. If simple averaging were performed, then the
outliers would skew the signal obtained. Algorithms such as 3$\sigma$ clipping that use outlier-sensitive dispersion measures
generally do not work either because these metrics are biased
by these fluctuations and define a very wide clipping range.
This is seen in Figure \ref{fig:sky}(d) (top plot), where the three frames
have been combined with an average after 3$\sigma$-clipping. Some of the pixels that exhibited S\&P also show deviating values in
the stacked frame.

It is common to use spatial filters for this kind of localized
noise. The most typical for S\&P is a median filter, where the
value of each pixel is replaced by the median of the values of
this and its close neighbors. This is done in the second row of
Figure \ref{fig:sky}(c). Note that the image has been smoothed and the
anomalous pixels may have disappeared. However, this has
been done at the expense of: (1) changing the signal and noise pattern of the image and (2) correlating the nearby pixels.
Although this is a very useful filter for improving the cosmetics
of the images, the photometry measurement in the resulting
image may be highly biased because the fluxes of each pixel
have been altered by its neighbors.\\

In this work, a solution based on convolutions and these two
previous ideas is proposed to try to mitigate the S\&P effect.
First, it should be considered that the S\&P effect mostly
impacts low noise areas, such as the sky background or faint
sources. For bright sources, the shot noise becomes higher than
the random telegraph noise and dominates all of the other
fluctuations. The median filter can be used to obtain a reference
frame to identify outliers because, in well-sampled fields, they
are a good approximation to a smoothed frame. This can be
seen in Figure \ref{fig:sky}(c), where the third row shows the difference
between the raw frame and the one filtered with a $3 \times 3$ median kernel. The residual pattern is generally homogeneous, both in
the sky area and in the vicinity of the sources. To identify the
S\&P, a threshold of 12 ADU has been set because it is at this
point that the distribution of three Gaussians in Figure \ref{fig:RON} (right-hand)
starts to be revealed. Hence, all of the pixels whose
absolute difference between the raw value and that resulting
from the convolution with the median filter is greater than 12
ADU are masked. In the bottom row of Figure \ref{fig:sky}(c), the raw
frames are shown with the pixels masked in red. By having a
sequence of frames, the average of the unmasked values of
each pixel can be taken to get a stacked image, which is shown
on the bottom right-hand of the figure, where the S\&P
contamination has been highly reduced. In cases where a pixel
shows S\&P in all of the frames of the sequence and is therefore
masked, the astronomer has to decide, for instance, either not to
take that pixel into account in the photometry or to replace its
value by an approximation, such as an interpolation of the
neighboring pixels or their median. In the example included
here, this happens in only two pixels out of $10^4$. It should be
noted that this method may be less accurate in fields with
critically sampled sources. A further review is currently
underway for future work.

Developing new algorithms or even using those already
available in common packages such as IRAF \citep{IRAF} or Astropy \citep{astropy} may not be trivial
when working with these cameras. The size of the raw 16 bit
images of the QHY600M Pro are about 120 MB, while those of
the QHY411M are 300 MB. If a sequence is taken at a high
frame rate, then the data set may not be manageable with
commonly available CPU capacities. The advantage of
convolution-based approaches and simple arithmetical operations,
such as median filtering or frame differencing, is that
they are easily deployable on GPUs, which allows the data to
be processed more efficiently and faster. The development of
GPU algorithms for astronomical image processing is essential
for further progress in the use of large sensors, such as these
sCMOS.

\section{Conclusion}
In the previous sections, the key features of the QHY600M
Pro and QHY411M cameras as scientific instruments have been
discussed in detail. For astronomy, they have characteristics
that make them very suitable for general use, although certain
issues need to be addressed. Our main conclusions are that:
\begin{enumerate}
    \item The built-in electronics in the pixels of sCMOS sensors
require that each pixel should be considered as an
individual detector and this should be taken into account
when performing processes such as bias or dark
subtraction. Spatial inhomogeneities in darkness are
detectable all over the frame.
    \item Their dark current is very low, as is the number of warm
pixels. They are stable for at least several weeks and their
signal scales linearly with exposure time, so they may be
quite easily removed with dark subtraction. This is an
improvement from previous sCMOS sensors and makes it
possible to take images with longer exposure times
without being affected by dark current.
    \item Its quantum efficiency peaks at 80\% at 475 nm and then drops rapidly at longer wavelengths, with 40\% at 700 nm and 10\% at 900 nm.
    \item They do not exhibit charge persistence or edge glow.
    \item These sensors are affected by random telegraph noise,
which can introduce non-negligible deviations in photometric
measurements of low-brightness sources. Simple
frame averaging, even with algorithms such as $\sigma$-clipping,
is not enough to mitigate its effect.
    \item They show promising performance on photometric
observations done with both fast and slow telescopes.
\end{enumerate}
Their low cost, power consumption, and replicability make
both cameras a very suitable solution for astronomical
applications, especially with regard to their high frame range,
near-zero readout time, and low readout noise. These sensors
are very good options for fast small telescopes with large fields
because of their small pixel size and large formats. The
combination of such telescopes and cameras permit very large
field-of-view images with plate scales with reasonably good
sampling of the PSF. For instance, a 11$^{\prime\prime}$ $f$/2.2 telescope such as
the Celestron RASA11 with a QHY600 camera can produce
images covering a FOV of 7.5 deg$^2$ with a plate scale of 1$^{\prime\prime}$.27 px$^{-1}$, which are excellent options for surveys such as
ATLAS-Teide \citep{Licandro2023}. Even so, having the
necessary computational tools to process the data, especially
GPU developments, is essential to take advantage of the full
performance of these cameras.\\

The authors declare no conflict of interest or relationship with the manufacturers of the cameras tested. M.R.A, M.S-R and J.L. acknowledge support from the ACIISI, Consejer\'{\i}a de Econom\'{\i}a, Conocimiento y Empleo del Gobierno de Canarias and the European Regional Development Fund (ERDF) under grant with reference ProID2021010134 and from the Agencia Estatal de Investigacion del Ministerio de Ciencia e I\'nnovacion (AEI-MCINN) under grant "Hydrated Minerals and Organic Compounds in Primitive Asteroids" with reference PID2020-120464GB-100. This research has been partially funded by Light Bridges, SL. which provided the QHY411M and Andor iKon-L 936 cameras for the tests presented here. This article includes observations made in the Two meter Twin Telescope (TTT) at the IAC's Teide Observatory that Light Bridges, SL, operates on the Island of Tenerife, Canary Islands (Spain).


\bibliography{PASPsample631}{}
\bibliographystyle{aasjournal}



\end{document}